\documentstyle[aps,epsfig]{revtex}
\begin{document}
\tightenlines
\title{Capture of $\alpha$ Particles by Isospin-Symmetric Nuclei}
\author{T. Rauscher, F.-K. Thielemann}
\address{Departement f\"ur Physik und Astronomie, Universit\"at
Basel, CH-4056 Basel, Switzerland}

\author{J. G\"orres, M. Wiescher}
\address{Department of Physics, University of Notre Dame, Notre
Dame, IN 46556, USA}

\maketitle
\begin{abstract}
The reaction rates for $\alpha$ capture processes on
self-conjugate nuclei in the mass range A=20-40 have been
investigated. The rates were calculated using the statistical model
code NON-SMOKER taking into account isospin suppression rules.
These theoretical predictions are compared with rates derived from
the available experimental data about the $\alpha$ capture
reactions but taking also into account additional experimental
information from different reaction channels populating the
$\alpha$ unbound states of the self-conjugate compound nuclei.
\vspace{0.5cm}

\end{abstract}

\normalsize
\section{Introduction}
The astrophysical importance of $\alpha$ capture on target nuclei
with $N=Z$ is manifold. In the Ne- and O-burning phase of massive
stars, alpha capture reaction sequences are initiated at $^{24}$Mg
and $^{28}$Si, respectively, and determine the abundance
distribution prior to the Si-burning phase~\cite{CLS98}.
Nucleosynthesis in explosive Ne and explosive O burning in type II
supernovae depend on reaction rates for $\alpha$ capture on
$^{20}$Ne to $^{36}$Ar~\cite{Arn95,Thie97}. While many of these
processes are in quasistatistical equilibrium~\cite{Thie97,ThH96,ThH99},
the reaction rates itself are important for a reliable description
of nucleosynthesis in the subsequent cooling phase. An $\alpha$
capture chain on such self-conjugate nuclei actually determines
the production of $^{44}$Ti \cite{hoff98}, which contributes to
the supernova light curve by the energy release from
its $\beta$ decay to $^{44}$Ca via $^{44}$Sc.
Charged-particle reaction networks have to consider $\alpha$-capture rates
in the conditions of an $\alpha$-rich freeze-out \cite{frei98}, or
for an extended rp-process for proper treatment of the $\gamma$ induced
$\alpha$ break-up of selfconjugate nuclei~\cite{WGI94,sch98}. Such calculations involve
also highly unstable nuclei, thus calling for a reliable
prediction of the respective reaction rates.

Due to isospin selection rules, $E1$ $\gamma$ transitions with
isospin $T=0 \rightarrow T=0$ are forbidden. Likewise, $M1$
transitions will be strongly suppressed. Because of isospin
conservation, only states with isospin $T=0$ can be populated by
$\alpha$ capture on $N=Z$ ($T=0$) targets. This leads to a strong
suppression of the $\gamma$ transitions in the compound nucleus
and thus of the ($\alpha$,$\gamma$) cross section of
self-conjugate nuclei~\cite{WaW69}.

Previous theoretical work used in astrophysical calculations
either neglected this isospin effect \cite{arn72,thi87} or
accounted for it only in a phenomenological way with arbitrary
suppression factors \cite{hol76,woo78,cow91,sch98}. In this work
we want to improve on the prediction of reaction rates on
isospin-symmetric targets and compare the theoretical values to
newly compiled available experimental information. In section
\ref{HF} the statistical model of nuclear reactions is introduced
and the method for obtaining isospin suppression factors as well
as the resulting theoretical cross sections and reaction rates are
presented. Section \ref{exp} reviews the available experimental
information and presents reaction rates newly derived from all
available experimental data covering different reaction channels.
A summary and conclusion will be given in Section
\ref{conc}.

\section{The Statistical Model}
\label{HF}

\subsection{Introduction}

The majority of nuclear reactions in astrophysics can be described
in the framework of the statistical model (compound nucleus
mechanism, Hauser-Feshbach approach, HF; see
e.g.~\cite{hau52,har77}), provided that the level density of the
compound nucleus is sufficiently large in the contributing energy
window \cite{rau97}. This description assumes that the reaction
proceeds via a compound nucleus which finally decays into the
reaction products. With a sufficiently high level density, average
cross sections
\begin{equation}
\sigma^{\rm HF}=\sigma_{\rm form} b_{\rm dec} = \sigma_{\rm form}
{\Gamma_{\rm final} \over \Gamma_{\rm tot}}
\end{equation}
can be calculated which can be factorized into a cross section
$\sigma_{\rm form}$ for the formation of the compound nucleus and a
branching ratio $b_{\rm dec}$. This branching ratio describes the probability of the decay
into the channel of interest relative to the total decay probability
into all possible exit channels. The partial widths $\Gamma$ as well as
$\sigma_{\rm form}$ are related to (averaged) transmission coefficients,
which comprise the central quantities in any HF calculation.

Many nuclear properties enter the computation of the transmission
coefficients:
mass differences (separation energies), optical potentials, GDR widths,
level densities. The transmission coefficients can be modified due to
pre-equilibrium effects which are included in width fluctuation
corrections \cite{tep74} (see also \cite{rau97}, and references therein)
and by isospin effects. It is in the description of the nuclear
properties where the various HF models differ.

In astrophysical applications usually different aspects are emphasized
than in pure nuclear physics investigations. Many of
the latter in this long and well established field were focused on
specific reactions. All or most "ingredients", like optical
potentials for
particle transmission coefficients, level densities, resonance
energies and widths of giant resonances for
predicting E1 and M1 $\gamma$-transitions, were deduced from
experiments.
As long as the statistical model prerequisites are met, this will
produce
highly accurate cross sections.
For the majority of nuclei in astrophysical applications such
information is not available. The real challenge is thus not the
well-established statistical model, but rather to provide all these
necessary ingredients
in as reliable a way as possible, also for nuclei where none of such
information is available.

\subsection{The NON-SMOKER Code}

For the calculations presented in this work we utilized the recently
developed statistical model code NON-SMOKER \cite{rau97a}. The current
status of the code is outlined in the following.

For neutrons and protons the optical potential of Jeukenne,
Lejeune \& Mahaux \cite{jeu77} is used, which is based on
microscopic infinite nuclear matter calculations for a given
density, applied with a local density approximation. It includes
corrections of the imaginary part \cite{fantoni81,mahaux82}. The
potential of McFadden \& Satchler \cite{mcf66} is used for
$\alpha$ particles, which is based on extensive data. Deformed
nuclei are treated by an effective spherical potential of equal
volume (see e.g. \cite{thi87,cow91}).

The level density treatment has been recently improved \cite{rau97}.
Additionally, experimental level information (excitation energies,
spins, parities) are included \cite{ENSDF}, as well as experimental
nuclear masses \cite{audi}.

The $\gamma$-transmission coefficients have to include the
dominant E1 and M1 $\gamma$ transitions. The smaller, less
important M1 transitions have usually been treated with the simple
single particle approach $T\propto E^3$ (see e.g.\ \cite{bla52}).
The E1 transitions are usually calculated on the basis of the
Lorentzian representation of the Giant Dipole Resonance (GDR).
Many microscopic and macroscopic models have been devoted to the
calculation of GDR energies and widths. An excellent fit to the
GDR energies is obtained with the hydrodynamic droplet model
\cite{mye77}. An improved microscopic-macroscopic approach is
used, based on dissipation and the coupling to quadrupole surface
vibrations \cite{thi83}; see also \cite{cow91}. Most recently it
was shown \cite{gor98} that the inclusion of ``soft mode'' or
``pygmy'' resonances
might have important consequences on the E1 transitions in
neutron-rich nuclei far off stability. The pygmy resonances could be
caused by a neutron skin which generates soft vibrational modes
\cite{isa92}. It is still under discussion whether such modes exist.
As the effect is negligible for nuclei close to
stability, we did not include it in the calculations presented
here.

\subsection{Inclusion of Isospin Effects}

The original Hauser-Feshbach equation \cite{hau52} implicitly assumes complete
isospin mixing but can be generalized to explicitly treat the
contributions of the dense background states with isospin
$T^<=T^{\rm g.s.}$ and the isobaric analog states with $T^>=T^<+1$
\cite{gri71,har77,sar82,har86}. In reality, compound nucleus
states do not have unique isospin and for that reason an isospin
mixing parameter $\mu\downarrow$ was introduced \cite{gri71},
which is the fraction of the width of $T^>$ states leading to
$T^<$ transitions; for complete isospin mixing $\mu\downarrow=1$,
for pure $T^<$ states $\mu\downarrow=0$. In the case of
overlapping resonances for each involved isospin, $\mu\downarrow$
is directly related to the level densities $\rho^<$ and $\rho^>$,
respectively. Isolated resonances can also be included via their
internal spreading width $\Gamma^{\downarrow}$ and a bridging
formula was derived to cover both regimes \cite{lan78}.

In order to determine the mixing parameter
$\mu\downarrow=\mu\downarrow(E)$, experimental information for
excitation energies of $T^>$ levels is used where available
\cite{ENSDF,rei90} in the code NON-SMOKER. Experimental values for
spreading widths are also tabulated \cite{har86,rei90}. Similarly
to the standard treatment for the $T^<$ states, a level density
description \cite{rau97} is invoked above the last experimentally
known $T^>$ level. Since the $T^>$ states in a nucleus ($Z$,$N$)
are part of a multiplet, they can be approximated by the levels
(and level density) of the nucleus ($Z$$-$1,$N$+1), only shifted
by a certain energy $E_{\rm d}$. This displacement energy $E_{\rm d}$ can be
calculated \cite{aue72} and it is dominated by the Coulomb
displacement energy: $E_{\rm d}=E_{\rm d}^{\rm Coul}+\epsilon$. In
the absence of experimental level information, we use the formula
from Woosley \& Fowler as given by \cite{ful82} for the
determination of the excitation energy of the first isobaric
analog state.

The inclusion of the explicit treatment of isospin has two major effects
on statistical cross section calculations in astrophysics:
the suppression of $\gamma$ widths for reactions involving
self-conjugate nuclei and the suppression of the neutron emission
in proton-induced reactions. This paper focuses on the suppression of
the $\gamma$ width in $\alpha$ capture reactions.
Non-statistical effects, i.e.\ the appearance of isobaric analog resonances,
will not be further discussed here.

The isospin selection rule for $E1$ transitions is $\Delta T=0,1$ with
transitions $0\rightarrow0$ being forbidden.
In the case of ($\alpha$,$\gamma$) reactions on targets with $N=Z$, the
cross sections will be heavily suppressed because $T=1$ states cannot be
populated in the compound nucleus due to isospin conservation.
A suppression will also be found
for capture reactions leading into self-conjugate nuclei, although
somewhat less pronounced because $T=1$ states can be populated according
to the isospin coupling coefficients.

In previous reaction rate calculations
\cite{hol76,woo78,cow91,sch98} the suppression of the $\gamma$
widths was treated completely phenomenologically by dividing the
{\it total} $\gamma$ widths (and thus the cross section) by quite
uncertain factors of 5 and 2, for ($\alpha$,$\gamma$) reactions
on self-conjugate nuclei
and nucleon capture reactions going into self-conjugate nuclei,
respectively, regardless of the
target's nature. These empirical factors were estimated from the
scarce experimental data~\cite{Coo77} available at that time.

We are replacing these factors by including more isospin
information into the calculation of the $E1$ and $M1$ suppression.
It can be shown (see e.g. \cite{jon87}) that $0\rightarrow0$ $E1$
transitions are forbidden. An approximate suppression rule for
$\Delta T=0$ transitions in self-conjugate nuclei can also be
derived for $M1$ transitions \cite{jon87} and leads to a
suppression factor of about 1/150. The total suppression of the
$E1$ transitions would be exact if isospin were an exact quantum
number, at least in the simplifying limit that the wavelength of
the transition involved is large compared with the nuclear size.
However, the Coulomb force mixes states of different isospin to a
small extent, so the ``clean'' selection rule becomes a
suppression factor, too. The theoretical estimate for the factor
as given by Ref.\ \cite{jon87} is 0.01.

Since only $T^<=0$ states can be populated by $\alpha$ capture,
we need
to know the mixing of these $T^<$ into $T^>$ states.
Below the first isobaric analog state at $E_{\rm d}$, states should be
unmixed.
We describe that phenomenologically by suppressing both $E1$ and $M1$
transitions. The suppression factor $f_{\rm iso}$
is set equal for $E1$ and $M1$.

It is clear that the total suppression factor of the resulting $\gamma$
width must be related to the number of resonances in the compound
nucleus. Therefore, in a straightforward way we set the suppression
for $\gamma$ transitions from the excitation energy $E$ of the compound nucleus
proportional to the ratio of
the density of $T=1$ levels and the density of $T=0$ levels,
\begin{equation}
f_{\rm iso}(E)\propto \frac{\rho^>(E)}{\rho^<(E)}\quad.
\end{equation}
We find a weak energy dependence of $f_{\rm iso}$.

\subsection{Results}

The cross sections and reaction rates
for $\alpha$ capture were calculated for the
$N=Z$ isotopes from Mg to Mo ($12 \leq Z \leq 42$).
The NON-SMOKER results including the above description of isospin
suppression of the $\gamma$ width are given in Tables \ref{T7} to \ref{T12}.
In the next section we compare these results to experimental cross
sections and rates, either directly measured or calculated from
resonance parameters.

\section{Comparison to experimental information}
\label{exp}

\subsection{Experimental data}
Experimental data on $\alpha$ capture reactions on self-conjugate
nuclei 20$\le$A$\le$40 are rather sparse. To determine reliably
the reaction rate at various stellar temperatures T$_9$ (in GK)
detailed information on the number of contributing resonances
$n$, the resonance strengths $\omega\gamma_i$ (in units eV), and
the resonance energies E$_i$ (in units MeV) are necessary:

\begin{equation}
N_A<\sigma v>~=~ 1.54\cdot 10^5\cdot A^{-3/2}
T_9^{-3/2}\cdot\sum_i^n\omega\gamma_i\cdot e^{\frac{-11.605\cdot
E_i}{T_9}} \quad ,
\end{equation}

\noindent with A as the reduced mass of the system. The resonance
strength depends on the spin of the resonance $J$ and the partial
widths of the entrance $\Gamma_{\alpha}$ and exit channel
$\Gamma_{\gamma}$, and the total width $\Gamma$

\begin{equation}
\omega\gamma~=~(2J+1)\cdot\frac{\Gamma_{\alpha}\cdot\Gamma_{\gamma}}{\Gamma}\quad.
\end{equation}

\noindent Extensive resonance studies in the astrophysically
relevant low energy range E$_{\alpha}\le$1.5~MeV have only been
made for $^{20}$Ne($\alpha$,$\gamma$)$^{24}$Mg \cite{sch83},
$^{24}$Mg($\alpha$,$\gamma$)$^{28}$Si \cite{maa78,CKM82}, and also for
$^{28}$Si($\alpha$,$\gamma$)$^{32}$S \cite{toe71,rog77}.
For the reaction $^{32}$S($\alpha$,$\gamma$)$^{36}$Ar only
insufficient data are available on resonances in the energy
range E$_{\alpha}\ge$2.2~MeV \cite{Ern64,Cla71,Cha83}. The
experimentally observed level density in the compound nucleus
$^{36}$Ar indicates a significantly higher number of possible
resonances \cite{end90}. Even less information is available on
resonances in $^{36}$Ar($\alpha$,$\gamma$)$^{40}$Ca. Resonance
measurements have been performed in the energy range of the giant
dipole resonance E$_{\alpha}$=6-17~MeV. Several strong resonances
have been observed and the resonance strengths were
determined~\cite{Wat73}. Measurements at lower energies, however,
E$_{\alpha}$=3-6~MeV, did not yield any significant information
on possible resonance states in this range~\cite{Nah67}. More
experimental data are again available for the reaction
$^{40}$Ca($\alpha$,$\gamma$)$^{44}$Ti. Several resonances were
successfully measured in the energy range E$_{\alpha}$=2.75-4~MeV
\cite{Coo77}, additional measurements were performed in the energy
range E$_{\alpha}$=4-6~MeV \cite{Dix77,Dix81}. Again, no
information is available on lower energy resonances.

Due to the lack of low energy resonance information a direct
comparison of the predicted HF rates and the rates
obtained by the experimental data is only of limited use. However,
additional and complementary information can be gained from
resonance $\alpha$ elastic scattering measurements and lifetime
measurements which yield information about the total width
$\Gamma$ of the resonance state. More important, however, are
(p,$\alpha)$ studies with a self-conjugate compound nucleus. These
measurements yield extensive information about the existence and
characteristics of possible resonances in the $\alpha$ capture
channel. In particular reactions like
$^{23}$Na(p,$\alpha$)$^{20}$Ne~\cite{end90,wgr89},
$^{27}$Al(p,$\alpha$)$^{24}$Mg~\cite{end90},
$^{31}$P(p,$\alpha$)$^{28}$Si~\cite{Ili91,Ili93,Ros95}
$^{35}$Cl(p,$\alpha$)$^{32}$S~\cite{Ili94,Ros95}, and
$^{39}$K(p,$\alpha$)$^{36}$Ar populate $\alpha$-unbound natural
parity states in the self-conjugate compound nuclei and therefore
complement the direct experimental information on resonances in
$^{20}$Ne($\alpha,\gamma$)$^{24}$Mg,
$^{24}$Mg($\alpha,\gamma$)$^{28}$Si,
$^{28}$Si($\alpha,\gamma$)$^{32}$S,
$^{32}$S($\alpha,\gamma$)$^{36}$Ar, and
$^{36}$Ar($\alpha,\gamma$)$^{40}$Ca, respectively. Particularly
useful is information on the resonance strength,
\begin{equation}
\omega\gamma_{(p,\alpha)}~=~\frac{2J+1}{(2j_p+1)(2j_t+1)}\cdot\frac{\Gamma_p\cdot\Gamma_{\alpha}}{\Gamma}
\end{equation}
with $j_p$ and $j_t$ as projectile and target spin. Combined with
information on the total width
\begin{equation}
\Gamma~\approx~\Gamma_p +\Gamma_{\alpha}
\end{equation}
and on
the resonance strengths of the correlated natural parity (p,$\gamma$) resonances
\begin{equation}
\omega\gamma_{(p,\gamma)}~=~\frac{2J+1}{(2j_p+1)(2j_t+1)}\cdot\frac{\Gamma_p\cdot\Gamma_{\gamma}}{\Gamma}.
\end{equation}
Such information allows to fit the partial widths $\Gamma_p$,
$\Gamma_{\gamma}$, $\Gamma_{\alpha}$ to match the total width and
the observed (p,$\gamma$) and (p,$\alpha$) strengths. The
($\alpha,\gamma$) resonance strength is calculated from these
values. If only partial data about (p,$\gamma$) or (p,$\alpha$)
resonance strengths and the total width is available, we adopted
an alpha spectroscopic factor from empirical alpha-strength
studies in self-conjugate nuclei in this mass range~\cite{Lev88}
and calculated the $\alpha$ width of the level in terms of a
simple potential model. If only either the total width, or one of
the (p,$\gamma$) or (p,$\alpha$) strengths is available, both the
alpha as well as the proton partial width need to be calculated.
For the latter case the single particle spectroscopic factor was
adopted from the average of the single particle strength distribution in the
excitation range of this nucleus. For natural parity levels with
no spectroscopic information available, the gamma width was
calculated additionally. We used an average Weisskopf strength
which has been matched to the known gamma strength distribution of
the neighboring states to account empirically for the isospin
$\gamma$ strength suppression in self-conjugate nuclei. Since
there are no data available for proton capture on the short-lived
$^{43}$Sc, the strengths of low energy resonances in
$^{40}$Ca($\alpha,\gamma$)$^{44}$Ti are based purely on such
estimates. The results are listed and compared with the available
experimental data in Table \ref{T1} to \ref{T6}. It can clearly be
seen that there is systematically good agreement between the
calculated and experimental resonance parameters.

\subsection{Reaction Rates}
The resonance parameters derived and discussed in the previous
chapter allow us to calculate the reaction rate N$_A<\sigma v>$ as a function
of temperature using Equation~2. These rates are directly
compared with the reaction rates N$_A<\sigma v>_{\rm HF}$ based on the
HF calculations with the code NON-SMOKER in
Tables \ref{T7} to \ref{T12}. Shown is the 'experimental' reaction rate
N$_A<\sigma v>_{\rm exp}$ which is based on the few directly observed
($\alpha,\gamma$) resonances only.
The 'empirical' rate
N$_A<\sigma v>_{\rm emp}$ is calculated using Equation~2 on the basis of
the extended resonance
set and the associated level parameter analysis discussed in the previous
section. The ratios of the 'experimental' to the 'empirical' rates is shown in
Figs.\ \ref{fig:ne20}$-$\ref{fig:ca40}.
For the reactions $^{20}$Ne($\alpha,\gamma$)$^{24}$Mg,
$^{24}$Mg($\alpha,\gamma$)$^{28}$Si, and
$^{28}$Si($\alpha,\gamma$)$^{21}$S
a considerable amount of experimental alpha capture data is available
for higher energies {\cite{sch83,maa78,toe71,rog77}, therefore at
temperatures above T$\approx$0.3~GK
the experimental rates are in good agreement with the empirical
rates, only at lower temperatures the experimental rates are
substantially lower due to the lack of experimental data.
Only very limited
data on alpha capture resonances are available for
$^{32}$S($\alpha,\gamma$)$^{36}$Ar and
$^{36}$Ar($\alpha,\gamma$)$^{40}$Ca, this explains the
substantial deviation between experimental and
empirical rate over the entire temperature range.

The ratios of the statistical model rates discussed in section \ref{HF} and empirical rates
discussed in section \ref{exp} are shown in Fig.\ \ref{fig:ratio} for the
astrophysically relevant temperature region.
Except for $^{32}$Si($\alpha,\gamma$)$^{36}$Ar and
$^{36}$Ar($\alpha,\gamma$)$^{40}$Ca at temperature T$< 2$~GK
the deviations remain
in the range of a factor of $\le$3. This is to be expected from a global
statistical model calculation. Towards higher temperature ($T\geq 3.5$ GK)
the calculation is significantly improved in respect to the empirical
rate and reaches a deviation of about 30\%.

Towards lower temperature two different behaviors can be identified.
The ratios of the theoretical and empirical rates for $^{20}$Ne($\alpha,\gamma$)$^{24}$Mg and
$^{40}$Ca($\alpha,\gamma$)$^{44}$Ti increase towards lower temperatures,
whereas the other ratios first decrease and finally strongly
increase at the lowest temperatures $T\leq 0.2$ GK
(not shown in Fig.\ \ref{fig:ratio} as the
corresponding rates are already very low and outside the temperature region
of interest).
In the latter very low temperature
range the reaction rates are
dominated by low energy resonances, E$_R\le$1~MeV. For these
resonances the $\alpha$ width $\Gamma_{\alpha}$ is typically
smaller than the $\gamma$ width and determines the resonance
strength $\omega\gamma_{(\alpha,\gamma)}$. The reduction of
$\gamma$-strength has no influence for these states. With only a few
dominating resonances the assumptions of the statistical model are not valid
anymore.

For the decrease of the ratios in the temperature range $T<3$ GK
the proper energy
dependence of the optical $\alpha$+nucleus potential is an important
factor. The slope of the ratio plotted in Fig.\ \ref{fig:ratio} for
each reaction is sensitive to the choice of the optical potential. The
use of an equivalent square well potential which neglects absorption in
the Coulomb barrier leads to a slightly less steep slope. Apparently,
all available $\alpha$+nucleus potentials in statistical model
calculations cannot account for the proper energy dependence at energies
close to the Coulomb barrier. This is a well-known problem in such
calculations.

The cases for $^{20}$Ne($\alpha,\gamma$)$^{24}$Mg and
$^{40}$Ca($\alpha,\gamma$)$^{44}$Ti seem different,
there the HF predictions are larger than the
empirical rate and the ratios increase monotonically towards lower
temperatures.
However, the HF calculation is in agreement within a factor of $\approx$2 for the
temperature range T$\ge$0.4~GK with the
experimental value directly derived from $\alpha$ capture resonances in
the reaction $^{20}$Ne($\alpha,\gamma$).
For $^{40}$Ca($\alpha,\gamma$), the empirical rate is nearly identical with the
experimental rate for temperatures T=1-5 GK as can be seen from figure \ref{fig:ca40}.
Most of the available experimental
information about natural parity states in $^{44}$Ti is based on
$\alpha$ capture studies~\cite{Coo77,Dix77,Dix81}. Only two
$\alpha$ unbound states at lower energies, E$_x\le$7.5~MeV, are
known from $^{46}$Ti(p,t)$^{44}$Ti two-particle transfer
measurements \cite{end90}. Also the results of
$^{40}$Ca($^6$Li,d)$^{44}$Ti and $^{40}$Ca($^7$Li,t)$^{44}$Ti
$\alpha$-transfer measurements \cite{SFZ74,FBL77} indicate a
fairly low level density in the excitation range of interest.
Shell model calculations support the experimental results that the
level density in the pf-shell nucleus $^{44}$Ti is
low~\cite{fis99}. The low level density  in $^{44}$Ti may limit
the applicability of the HF approach for calculating
the reaction rate of $^{40}$Ca($\alpha,\gamma$)$^{44}$Ti.
This is supported by the fact that a variation in the level density
description employed in the statistical model calculation has a
significant impact on the reaction rate of
$^{40}$Ca($\alpha,\gamma$)$^{44}$Ti but leaves the remaining rates
nearly unchanged. This high sensitivity is due to the low level density
in the compound nucleus.

\section{Summary and Conclusion}
\label{conc}
In this paper we attempted to improve the statistical model
description of $\alpha$ capture rates on self-conjugate nuclei, which
are important for many nucleosynthesis processes in stellar and
explosive He-burning.

The NON-SMOKER predictions presented here are typically lower than the
results of previous Hauser Feshbach calculations~\cite{hol76,woo78}, except for
$^{28}$Si($\alpha$,$\gamma$)$^{32}$S for which the present theoretical
rate is higher by a factor of 1.96 (see also Table 6 in~\cite{hoff98}).
Those previous calculations approximated the isospin effect by simply
dividing the total $\gamma$ width by a factor of 5 for isospin
conjugated nuclei and employed equivalent square well potentials for the
calculation of the $\alpha$ transmission coefficients.
As expected, differences are larger in comparison to calculations
neglecting the isospin suppression.

The statistical model rates are compared with reaction rates derived
directly from experiment or calculated
from resonance parameters which have been measured through different
reaction channels.
These rates agree reasonably well with the HF
predictions in the astrophysically interesting temperature range $1<T_9<5$
(see Fig.\ \ref{fig:ratio}). At lower
temperatures, the differences tend to be
larger, due to the increasing level spacing and the importance of single
resonances.

As the isobaric energy shifts are quite different for the
considered compound nuclei but the general trend in the isospin suppression of
the $\gamma$ widths is nevertheless well reproduced in the theoretical
framework, we conclude that the approach presented here is valid in the energy
range of astrophysical importance. The remaining differences have to be
attributed to the description of the energy dependence of other nuclear
properties, such as the optical $\alpha$+nucleus potential or the level
density.

\acknowledgments
This work was partly supported by the Swiss NSF (grant 2000-053798.98)
and by the National Science Foundation (grant PHY98-03757).
T. R. is a PROFIL fellow of the Swiss NSF (grant 2124-055832.98).

\clearpage
\begin{table}
\caption{\label{T1} Resonance parameters for the
$^{20}$Ne($\alpha,\gamma$)$^{24}$Mg reaction.}
\begin{tabular}{rcclll}
E$_x$ & E$_r^{cm}$ & J$^{\pi}$ & $\Gamma_{tot}$ &
$\omega\gamma_{calc}$ & $\omega\gamma_{exp}$ \\
\hline
[MeV] & [MeV] &  & [eV] & [eV] & [eV]\\
\hline
10.111 &  0.7992 &  0$^+$ &  3.60E-01  &  2.86E-04  &   2.90E-04   \\
10.161 &  0.8492 &  0$^+$ &  3.68E-01  &  2.30E-05  &              \\
10.679 &  1.3672 &  0$^+$ &  2.29E+00  &  1.66E-01  &   1.70E-01   \\
11.457 &  2.1452 &  0$^+$ &  1.01E+03  &  2.00E-02  &   2.00E-02   \\
11.727 &  2.4152 &  0$^+$ &  1.00E+04  &  3.70E-01  &   3.70E-01   \\
13.04  &  3.7282 &  0$^+$ &  2.30E+03  &  6.71E-01  &   7.30E-01   \\
13.198 &  3.8862 &  0$^+$ &  2.70E+03  &  4.83E-01  &              \\
       &         &        &            &            &              \\
10.362 &  1.0502 &  1$^-$ &  4.06E-01  &  4.91E-04  &   4.80E-04   \\
10.659 &  1.3472 &  1$^-$ &  4.93E-01  &  7.25E-02  &              \\
11.39  &  2.0782 &  1$^-$ &  5.00E+02  &  4.60E-01  &   4.60E-01   \\
11.864 &  2.5522 &  1$^-$ &  7.00E+03  &  2.40E+00  &   1.00E+00   \\
12.447 &  3.1352 &  1$^-$ &  5.56E+03  &  3.97E-05  &              \\
13.253 &  3.9412 &  1$^-$ &  3.60E+04  &  4.79E-01  &              \\
13.273 &  3.9612 &  1$^-$ &  2.00E+03  &  1.67E-01  &              \\
13.332 &  4.0202 &  1$^-$ &  3.10E+04  &  1.21E-01  &              \\
13.407 &  4.0952 &  1$^-$ &  2.80E+03  &  2.90E+00  &   2.90E+00   \\
13.581 &  4.2692 &  1$^-$ &  2.10E+04  &  2.50E-01  &              \\
13.585 &  4.2732 &  1$^-$ &  8.00E+03  &  6.60E-01  &              \\
13.81  &  4.4982 &  1$^-$ &  2.40E+04  &  2.87E-01  &              \\
14.019 &  4.7072 &  1$^-$ &  6.20E+03  &  1.25E+00  &   1.20E+01   \\
       &         &        &            &            &              \\
9.4578 &  0.146  &  2$^+$ &  1.10E-01  &  1.86E-28  &              \\
10.361 &  1.0489 &  2$^+$ &  5.16E-02  &  7.52E-03  &              \\
10.731 &  1.4192 &  2$^+$ &  4.97E-03  &  5.08E-03  &              \\
10.917 &  1.6052 &  2$^+$ &  7.34E+00  &  2.45E+00  &   2.20E+00   \\
11.016 &  1.7042 &  2$^+$ &  1.19E+00  &  1.48E+00  &   1.50E+00   \\
11.208 &  1.8962 &  2$^+$ &  2.56E-03  &  2.74E-03  &              \\
11.293 &  1.9812 &  2$^+$ &  1.93E-04  &  6.18E-05  &              \\
11.33  &  2.0182 &  2$^+$ &  1.71E+01  &  3.06E+00  &              \\
11.453 &  2.1412 &  2$^+$ &  3.24E+01  &  3.28E+00  &   1.20E+00   \\
11.519 &  2.2072 &  2$^+$ &  5.02E+02  &  7.10E-01  &   8.00E-01   \\
11.964 &  2.6522 &  2$^+$ &  2.35E+03  &  8.58E-01  &   8.50E-01   \\
11.988 &  2.6762 &  2$^+$ &  7.61E-01  &  9.48E-03  &              \\
12.181 &  2.8692 &  2$^+$ &  1.73E+02  &  1.17E+00  &   9.00E-01   \\
12.403 &  3.0912 &  2$^+$ &  3.90E+02  &  6.09E-02  &              \\
12.467 &  3.1552 &  2$^+$ &  3.84E+03  &  2.11E+00  &   2.30E+00   \\
12.577 &  3.2652 &  2$^+$ &  6.43E+03  &  1.08E+00  &   9.00E-01   \\
12.737 &  3.4252 &  2$^+$ &  7.81E+03  &  1.64E+00  &   7.00E-01   \\
12.776 &  3.4642 &  2$^+$ &  3.11E+04  &  8.67E-03  &              \\
12.805 &  3.4932 &  2$^+$ &  1.09E+03  &  2.10E+00  &   1.80E+00   \\
12.845 &  3.5332 &  2$^+$ &  2.00E+02  &  2.02E-01  &              \\
13.03  &  3.7182 &  2$^+$ &  1.80E+03  &  2.85E-03  &              \\
13.08  &  3.7682 &  2$^+$ &  1.19E+04  &  2.05E+00  &   2.00E+00   \\
13.137 &  3.8252 &  2$^+$ &  5.43E+03  &  6.60E+00  &              \\
13.417 &  4.1052 &  2$^+$ &  4.50E+03  &  5.58E-03  &              \\
13.451 &  4.1392 &  2$^+$ &  2.10E+03  &  6.19E-02  &              \\
13.473 &  4.1612 &  2$^+$ &  1.00E+03  &  2.24E-03  &              \\
13.675 &  4.3632 &  2$^+$ &  5.40E+03  &  3.83E-01  &              \\
13.72  &  4.4082 &  2$^+$ &  2.74E+03  &  1.13E+00  &              \\
13.884 &  4.5722 &  2$^+$ &  4.80E+04  &  1.70E+00  &              \\
14.024 &  4.7122 &  2$^+$ &  7.00E+03  &  2.66E-03  &              \\
14.099 &  4.7872 &  2$^+$ &  1.40E+03  &  1.30E+00  &   1.30E+00   \\
       &         &        &            &            &              \\
9.532  &  0.2202 &  3$^-$ &  2.68E-01  &  1.42E-22 &               \\
10.333 &  1.0212 &  3$^-$ &  4.01E-01  &  3.21E-04 &    3.00E-04   \\
10.659 &  1.3472 &  3$^-$ &  4.69E-01  &  1.08E-02 &               \\
11.162 &  1.8502 &  3$^-$ &  6.21E-01  &  2.12E-01 &    2.20E-01   \\
11.595 &  2.2832 &  3$^-$ &  1.60E-01  &  3.55E-02 &    3.70E-02   \\
11.998 &  2.6862 &  3$^-$ &  8.68E+00  &  3.41E-01 &    3.75E-01   \\
12.015 &  2.7032 &  3$^-$ &  7.15E+02  &  5.95E+00 &               \\
12.257 &  2.9452 &  3$^-$ &  1.23E+03  &  6.32E+00 &               \\
12.659 &  3.3472 &  3$^-$ &  9.01E+02  &  5.05E+00 &               \\
12.919 &  3.6072 &  3$^-$ &  6.67E+03  &  7.19E-01 &               \\
13.087 &  3.7752 &  3$^-$ &  8.96E+03  &  2.54E-01 &               \\
13.423 &  4.1112 &  3$^-$ &  4.51E+03  &  5.14E-01 &               \\
13.445 &  4.1332 &  3$^-$ &  2.00E+02  &  1.05E-01 &               \\
       &         &        &    &   &                       \\
10.581 &  1.2692 &  4$^+$ &  4.51E-01  &  6.37E-04 &               \\
10.66  &  1.3482 &  4$^+$ &  4.68E-01  &  1.67E-03 &               \\
11.217 &  1.9052 &  4$^+$ &  8.72E-01  &  1.67E+00 &    1.70E+00   \\
11.314 &  2.0022 &  4$^+$ &  6.81E-01  &  4.25E-01 &               \\
11.695 &  2.3832 &  4$^+$ &  1.60E+00  &  1.12E+00 &    1.10E+00   \\
12.049 &  2.7372 &  4$^+$ &  2.74E-01  &  4.68E-01 &               \\
12.117 &  2.8052 &  4$^+$ &  1.90E+03  &  1.41E+00 &    1.40E+00   \\
12.161 &  2.8492 &  4$^+$ &  9.00E+02  &  7.18E-01 &               \\
12.504 &  3.1922 &  4$^+$ &  1.73E+02  &  6.15E+00 &    6.20E+00   \\
12.636 &  3.3242 &  4$^+$ &  3.00E+01  &  8.67E-01 &               \\
12.972 &  3.6602 &  4$^+$ &  3.30E+03  &  9.67E-02 &               \\
13.05  &  3.7382 &  4$^+$ &  9.06E+01  &  5.78E+00 &               \\
14.148 &  4.8362 &  4$^+$ &  1.80E+03  &  1.25E+00 &    1.50E+00   \\
14.327 &  5.0152 &  4$^+$ &  9.02E+03  &  6.68E+00 &    6.40E+00   \\
\end{tabular}
\end{table}
\newpage
\begin{table}
\caption{\label{T2} Resonance parameters for the
$^{24}$Mg($\alpha,\gamma$)$^{28}$Si reaction.}
\begin{tabular}{rcclll}
E$_x$ & E$_r^{cm}$ & J$^{\pi}$ & $\Gamma_{tot}$ &
$\omega\gamma_{calc}$ & $\omega\gamma_{exp}$ \\
\hline
[MeV] & [MeV] &  & [eV] & [eV] & [eV]\\
\hline
12.976 &  2.992 &  0$^+$ & 5.20E+03 & 5.25E-02 &             \\
13.039 &  3.055 &  0$^+$ & 3.20E+03 & 1.77E-01 &   3.00E-01  \\
13.235 &  3.251 &  0$^+$ & 3.00E+03 & 3.53E-01 &   1.30E+00  \\
       &        &        &          &          &             \\
11.295 &  1.311 &  1$^-$ & 9.50E-02 & 9.08E-03 &   1.10E-01  \\
11.669 &  1.685 &  1$^-$ & 4.60E-01 & 2.13E-01 &   3.30E-01  \\
12.181 &  2.197 &  1$^-$ & 8.26E+00 & 9.47E-02 &   1.00E-01  \\
12.3   &  2.316 &  1$^-$ & 2.20E+01 & 2.19E-02 &   2.00E-02  \\
12.815 &  2.831 &  1$^-$ & 3.50E+03 & 5.17E-01 &   2.00E-01  \\
12.973 &  2.989 &  1$^-$ & 1.69E+03 & 1.77E+01 &   8.00E-01  \\
13.423 &  3.439 &  1$^-$ & 1.97E+04 & 1.72E+01 &             \\
13.734 &  3.75  &  1$^-$ & 3.50E+04 & 4.77E-02 &             \\
13.812 &  3.828 &  1$^-$ & 3.70E+03 & 4.83E-01 &   3.00E-01  \\
13.897 &  3.913 &  1$^-$ & 5.40E+03 & 5.28E-01 &             \\
13.986 &  4.002 &  1$^-$ & 2.71E+03 & 2.18E-02 &             \\
       &        &        &          &          &             \\
10.209 &  0.225 &  2$^+$ & 3.39E-02 & 2.17E-26 &             \\
10.514 &  0.53  &  2$^+$ & 6.42E-02 & 2.18E-12 &             \\
10.805 &  0.821 &  2$^+$ & 7.36E-02 & 2.74E-07 &             \\
10.915 &  0.931 &  2$^+$ & 7.75E-02 & 4.99E-06 &             \\
10.994 &  1.01  &  2$^+$ & 3.13E-02 & 2.95E-05 &             \\
10.951 &  0.967 &  2$^+$ & 7.88E-02 & 1.15E-05 &             \\
11.078 &  1.094 &  2$^+$ & 8.35E-02 & 1.56E-04 &             \\
11.138 &  1.154 &  2$^+$ & 8.58E-02 & 4.58E-04 &             \\
11.265 &  1.281 &  2$^+$ & 9.14E-02 & 3.39E-03 &             \\
11.432 &  1.448 &  2$^+$ & 1.04E-01 & 2.94E-02 &             \\
11.515 &  1.531 &  2$^+$ & 1.17E-01 & 6.99E-02 &   6.00E-02  \\
11.656 &  1.672 &  2$^+$ & 1.80E-01 & 2.13E-01 &   1.40E-01  \\
11.778 &  1.794 &  2$^+$ & 2.16E-01 & 2.72E-02 &   3.00E-02  \\
12.071 &  2.087 &  2$^+$ & 2.18E+00 & 5.51E-01 &   3.60E-01  \\
12.289 &  2.305 &  2$^+$ & 2.10E+01 & 1.01E-01 &   9.00E-02  \\
12.44  &  2.456 &  2$^+$ & 1.76E+01 & 1.70E+00 &   1.00E+00  \\
12.714 &  2.73  &  2$^+$ & 5.01E+01 & 2.14E-04 &             \\
12.725 &  2.741 &  2$^+$ & 6.43E+02 & 1.27E+00 &   3.60E+00  \\
12.754 &  2.77  &  2$^+$ & 5.07E+01 & 6.02E-02 &             \\
12.899 &  2.915 &  2$^+$ & 9.59E+02 & 5.20E+00 &   2.40E+00  \\
12.923 &  2.939 &  2$^+$ & 2.31E+02 & 5.58E-01 &   4.00E-01  \\
13.105 &  3.121 &  2$^+$ & 2.47E+02 & 2.84E+00 &   2.10E+00  \\
13.187 &  3.203 &  2$^+$ & 1.90E+03 & 1.85E-02 &             \\
13.203 &  3.219 &  2$^+$ & 2.14E+02 & 1.23E-01 &             \\
13.228 &  3.244 &  2$^+$ & 2.25E+02 & 1.12E+00 &   1.30E+00  \\
13.42  &  3.436 &  2$^+$ & 1.12E+03 & 4.32E-01 &   3.70E-01  \\
13.461 &  3.477 &  2$^+$ & 1.80E+03 & 3.90E-01 &             \\
13.482 &  3.498 &  2$^+$ & 1.53E+03 & 1.03E-01 &             \\
13.545 &  3.561 &  2$^+$ & 8.47E+03 & 2.71E-02 &             \\
13.638 &  3.654 &  2$^+$ & 4.50E+03 & 5.12E-02 &   9.00E-02  \\
13.639 &  3.655 &  2$^+$ & 1.20E+02 & 6.68E-01 &             \\
13.677 &  3.693 &  2$^+$ & 1.37E+03 & 1.41E+00 &   1.10E+00  \\
13.705 &  3.721 &  2$^+$ & 5.00E+02 & 1.66E+00 &             \\
13.939 &  3.955 &  2$^+$ & 5.17E+03 & 3.93E+00 &   6.00E-02  \\
13.971 &  3.987 &  2$^+$ & 2.52E+03 & 1.86E+00 &   6.00E-01  \\
13.983 &  3.999 &  2$^+$ & 3.84E+02 & 1.80E+01 &             \\
14.064 &  4.08  &  2$^+$ & 6.00E+03 & 5.77E-01 &   8.00E-01  \\
14.309 &  4.325 &  2$^+$ & 2.00E+04 & 5.11E+00 &   2.00E+00  \\
       &        &        &          &          &             \\
10.181 &  0.197 &  3$^-$ & 1.00E-01 & 8.12E-30 &             \\
10.54  &  0.556 &  3$^-$ & 6.50E-02 & 2.17E-12 &             \\
11.584 &  1.6   &  3$^-$ & 1.10E-01 & 3.93E-02 &   4.00E-02  \\
11.899 &  1.915 &  3$^-$ & 1.00E+02 & 6.37E-02 &   6.00E-03  \\
11.931 &  1.947 &  3$^-$ & 3.27E-01 & 4.32E-01 &             \\
11.975 &  1.991 &  3$^-$ & 7.00E-01 & 1.43E-01 &   1.00E-01  \\
12.134 &  2.15  &  3$^-$ & 1.00E+01 & 6.48E-02 &             \\
12.193 &  2.209 &  3$^-$ & 2.32E+01 & 1.56E-01 &   1.90E-01  \\
12.488 &  2.504 &  3$^-$ & 1.09E+02 & 1.88E-01 &   2.00E-01  \\
12.741 &  2.757 &  3$^-$ & 2.00E+03 & 1.87E-03 &             \\
12.801 &  2.817 &  3$^-$ & 9.72E+01 & 8.60E-03 &             \\
12.858 &  2.874 &  3$^-$ & 2.00E+02 & 5.33E-01 &   6.00E-01  \\
12.989 &  3.005 &  3$^-$ & 2.30E+03 & 1.28E-02 &             \\
13.114 &  3.13  &  3$^-$ & 1.52E+04 & 6.70E-03 &             \\
13.172 &  3.188 &  3$^-$ & 3.06E+02 & 9.01E-02 &             \\
13.246 &  3.262 &  3$^-$ & 9.97E+03 & 2.16E-01 &   2.00E-01  \\
13.317 &  3.333 &  3$^-$ & 1.20E+03 & 1.77E-01 &             \\
13.358 &  3.374 &  3$^-$ & 4.72E+03 & 2.07E-01 &   2.30E-01  \\
13.491 &  3.507 &  3$^-$ & 3.20E+04 & 4.93E-02 &             \\
13.662 &  3.678 &  3$^-$ & 4.50E+02 & 3.62E-01 &             \\
13.711 &  3.727 &  3$^-$ & 2.00E+04 & 2.03E-01 &   8.00E-01  \\
13.788 &  3.804 &  3$^-$ & 2.70E+03 & 3.32E-01 &             \\
13.835 &  3.851 &  3$^-$ & 9.00E+02 & 3.39E-01 &   1.00E-01   \\
13.859 &  3.875 &  3$^-$ & 3.87E+03 & 3.27E+00 &              \\
13.872 &  3.888 &  3$^-$ & 7.10E+03 & 9.05E-01 &   8.00E-01   \\
\end{tabular}
\end{table}
\newpage
\begin{table}
\caption{\label{T3} Resonance parameters for the
$^{28}$Si($\alpha,\gamma$)$^{32}$S reaction.}
\begin{tabular}{rcclll}
E$_x$ & E$_r^{cm}$ & J$^{\pi}$ & $\Gamma_{tot}$ &
$\omega\gamma_{calc}$ & $\omega\gamma_{exp}$ \\
\hline
[MeV] & [MeV] &  & [eV] & [eV] & [eV]\\
\hline
8.507 &  1.558 &  0$^+$ & 2.25E-02 & 2.61E-04  &           \\
9.983 &  3.034 &  0$^+$ & 1.19E+02 & 1.11E-01  &           \\
10.457&  3.508 &  0$^+$ & 1.73E+03 & 1.72E-02  &           \\
10.787&  3.838 &  0$^+$ & 6.00E+02 & 4.50E+01  &           \\
11.064&  4.115 &  0$^+$ & 1.25E+04 & 1.34E-03  &           \\
11.584&  4.635 &  0$^+$ & 3.19E+03 & 7.66E-02  &           \\
11.607&  4.658 &  0$^+$ & 3.10E+02 & 8.52E+00  &           \\
11.869&  4.92  &  0$^+$ & 1.10E+03 & 7.13E-02  &           \\
11.93 &  4.981 &  0$^+$ & 1.00E+02 & 4.93E-01  &           \\
      &        &        &          &           &           \\
7.434 &  0.485 &  1$^-$ & 1.14E-02 & 5.03E-18  &           \\
8.494 &  1.545 &  1$^-$ & 2.91E-02 & 1.59E-02  &  1.60E-02 \\
9.236 &  2.287 &  1$^-$ & 8.99E-01 & 5.94E-01  &  5.40E-01 \\
9.486 &  2.537 &  1$^-$ & 9.98E+00 & 4.48E-02  &  8.30E-01 \\
9.731 &  2.782 &  1$^-$ & 3.74E+02 & 1.85E-03  &           \\
9.849 &  2.9   &  1$^-$ & 1.02E+02 & 8.99E-03  &           \\
9.95  &  3.001 &  1$^-$ & 1.50E+02 & 5.48E-02  &           \\
10.226&  3.277 &  1$^-$ & 1.80E+02 & 1.48E-02  &           \\
10.332&  3.383 &  1$^-$ & 6.64E+03 & 4.82E-02  &           \\
10.604&  3.655 &  1$^-$ & 1.50E+02 & 2.55E-01  &           \\
10.701&  3.752 &  1$^-$ & 2.10E+04 & 2.23E-01  &           \\
10.711&  3.762 &  1$^-$ & 2.00E+04 & 3.21E+00  &  2.70E+00 \\
10.826&  3.877 &  1$^-$ & 2.26E+04 & 2.38E-02  &           \\
10.916&  3.967 &  1$^-$ & 1.63E+03 & 1.56E-01  &           \\
11.234&  4.285 &  1$^-$ & 7.90E+03 & 1.42E-02  &           \\
11.447&  4.498 &  1$^-$ & 5.30E+03 & 1.00E-02  &           \\
11.587&  4.638 &  1$^-$ & 1.64E+03 & 2.91E-01  &           \\
11.609&  4.66  &  1$^-$ & 1.93E+04 & 2.13E-02  &           \\
11.63 &  4.681 &  1$^-$ & 2.71E+04 & 2.09E-02  &           \\
11.735&  4.786 &  1$^-$ & 2.43E+04 & 1.81E-02  &           \\
11.807&  4.858 &  1$^-$ & 3.71E+04 & 3.43E-02  &           \\
11.91 &  4.961 &  1$^-$ & 6.30E+03 & 4.32E-01  &           \\
12.185&  5.236 &  1$^-$ & 1.07E+04 & 5.62E-02  &           \\
12.195&  5.246 &  1$^-$ & 5.63E+03 & 3.60E-02  &           \\
12.296&  5.347 &  1$^-$ & 1.56E+04 & 2.21E-02  &           \\
12.336&  5.387 &  1$^-$ & 2.80E+03 & 8.08E-03  &           \\
12.39 &  5.441 &  1$^-$ & 4.20E+03 & 1.63E-03  &           \\
      &        &        &          &           &           \\
7.484 &  0.535 &  2$^+$ & 1.17E-02 & 9.52E-17  &           \\
8.344 &  1.395 &  2$^+$ & 2.02E-02 & 3.20E-05  &           \\
8.69  &  1.741 &  2$^+$ & 2.74E-02 & 1.17E-02  &  1.20E-02 \\
8.861 &  1.912 &  2$^+$ & 3.15E-02 & 1.82E-02  &  1.60E-02 \\
9.196 &  2.247 &  2$^+$ & 4.45E-01 & 1.52E-01  &           \\
9.464 &  2.515 &  2$^+$ & 8.59E-01 & 7.12E-01  &  7.20E-01 \\
9.65  &  2.701 &  2$^+$ & 7.46E+00 & 9.53E-02  &           \\
9.711 &  2.762 &  2$^+$ & 4.68E+00 & 2.91E-01  &  6.30E-01 \\
9.817 &  2.868 &  2$^+$ & 6.82E+00 & 6.21E-03  &           \\
9.92  &  2.971 &  2$^+$ & 1.00E+01 & 3.68E-02  &           \\
10.293&  3.344 &  2$^+$ & 7.00E+01 & 2.52E-02  &           \\
10.51 &  3.561 &  2$^+$ & 1.00E+01 & 1.66E-02  &           \\
10.528&  3.579 &  2$^+$ & 8.01E+01 & 7.41E-02  &  7.00E-02 \\
10.696&  3.747 &  2$^+$ & 1.80E+02 & 1.01E-01  &           \\
10.757&  3.808 &  2$^+$ & 5.00E+01 & 1.32E-01  &           \\
10.779&  3.83  &  2$^+$ & 6.00E+02 & 3.69E-01  &           \\
10.792&  3.843 &  2$^+$ & 1.76E+02 & 1.33E+00  &           \\
10.827&  3.878 &  2$^+$ & 3.21E+02 & 5.35E+00  &           \\
10.933&  3.984 &  2$^+$ & 4.80E+01 & 3.57E-01  &           \\
11.051&  4.102 &  2$^+$ & 4.22E+03 & 1.66E-02  &           \\
11.083&  4.134 &  2$^+$ & 8.50E+01 & 2.11E-01  &           \\
11.237&  4.288 &  2$^+$ & 5.00E+01 & 2.12E-01  &           \\
11.255&  4.306 &  2$^+$ & 2.10E+02 & 2.64E-01  &           \\
11.333&  4.384 &  2$^+$ & 1.50E+02 & 2.23E-01  &           \\
11.487&  4.538 &  2$^+$ & 5.04E+02 & 6.57E-02  &           \\
11.604&  4.655 &  2$^+$ & 8.10E+02 & 6.73E-03  &           \\
11.679&  4.73  &  2$^+$ & 2.60E+03 & 2.50E-02  &           \\
11.722&  4.773 &  2$^+$ & 2.81E+03 & 1.00E-02  &           \\
11.832&  4.883 &  2$^+$ & 1.40E+02 & 1.35E-01  &           \\
12.00 &  5.051 &  2$^+$ & 2.00E+03 & 3.69E-03  &           \\
12.014&  5.065 &  2$^+$ & 9.01E+03 & 1.52E-01  &           \\
12.022&  5.073 &  2$^+$ & 2.63E+03 & 4.93E-02  &           \\
12.147&  5.198 &  2$^+$ & 1.32E+04 & 4.93E-02  &           \\
12.298&  5.349 &  2$^+$ & 3.26E+02 & 6.27E-01  &           \\
12.389&  5.44  &  2$^+$ & 1.85E+03 & 1.57E-01  &           \\
12.399&  5.45  &  2$^+$ & 3.61E+02 & 3.28E-01  &           \\
12.473&  5.524 &  2$^+$ & 1.40E+03 & 5.95E-03  &           \\
      &        &        &          &           &           \\
7.701 &  0.752 &  3$^-$ & 1.35E-02 & 1.64E-12  &           \\
9.023 &  2.074 &  3$^-$ & 4.02E-02 & 5.36E-02  &  5.20E-02 \\
9.464 &  2.515 &  3$^-$ & 1.07E-01 & 1.61E-01  &           \\
9.724 &  2.775 &  3$^-$ & 1.59E+00 & 1.57E-02  &           \\
9.817 &  2.868 &  3$^-$ & 4.15E+00 & 7.33E-03  &           \\
10.223&  3.274 &  3$^-$ & 6.36E+01 & 7.65E+00  &  8.10E+00 \\
10.288&  3.339 &  3$^-$ & 1.62E+02 & 2.19E+00  &  2.30E+00 \\
10.626&  3.677 &  3$^-$ & 6.68E+02 & 4.65E-01  &  6.00E-01 \\
11.092&  4.143 &  3$^-$ & 7.00E+01 & 2.93E-01  &           \\
11.197&  4.248 &  3$^-$ & 8.00E+01 & 3.44E-01  &           \\
11.231&  4.282 &  3$^-$ & 8.00E+02 & 3.78E-02  &           \\
11.557&  4.608 &  3$^-$ & 5.00E+01 & 2.74E-01  &           \\
11.625&  4.676 &  3$^-$ & 6.00E+01 & 2.68E-01  &           \\
11.758&  4.809 &  3$^-$ & 1.40E+02 & 1.54E-01  &           \\
11.784&  4.835 &  3$^-$ & 3.00E+01 & 7.60E-01  &           \\
11.805&  4.856 &  3$^-$ & 2.02E+03 & 7.98E-03  &           \\
11.879&  4.93  &  3$^-$ & 3.60E+03 & 3.86E-02  &           \\
11.902&  4.953 &  3$^-$ & 2.45E+03 & 3.74E-02  &           \\
11.936&  4.987 &  3$^-$ & 1.48E+03 & 4.73E-02  &           \\
12.143&  5.194 &  3$^-$ & 3.20E+03 & 1.59E-03  &           \\
12.154&  5.205 &  3$^-$ & 4.45E+03 & 5.84E-02  &           \\
12.272&  5.323 &  3$^-$ & 3.08E+02 & 8.98E-01  &           \\
12.421&  5.472 &  3$^-$ & 2.85E+03 & 3.28E-01  &           \\
12.457&  5.508 &  3$^-$ & 3.38E+03 & 1.41E-02  &           \\
\end{tabular}
\end{table}
\newpage
\begin{table}
\caption{\label{T4} Resonance parameters for the
$^{32}$S($\alpha,\gamma$)$^{36}$Ar reaction.}
\begin{tabular}{rcclll}
E$_x$ & E$_r^{cm}$ & J$^{\pi}$ & $\Gamma_{tot}$ &
$\omega\gamma_{calc}$ & $\omega\gamma_{exp}$ \\
\hline
[MeV] & [MeV] &  & [eV] & [eV] & [eV]\\
\hline
7.136 & 0.495 &  1$^-$ &  5.00E-02 & 4.30E-21 &            \\
7.247 & 0.606 &  1$^-$ &  3.00E-02 & 1.43E-17 &            \\
7.749 & 1.108 &  1$^-$ &  1.62E-02 & 4.83E-09 &            \\
7.879 & 1.238 &  1$^-$ &  1.76E-02 & 9.37E-08 &            \\
8.302 & 1.661 &  1$^-$ &  2.29E-02 & 2.06E-04 &            \\
9.117 & 2.476 &  1$^-$ &  1.16E+00 & 2.73E-01 &   3.00E-01 \\
9.366 & 2.725 &  1$^-$ &  1.38E+01 & 2.38E-01 &            \\
9.702 & 3.061 &  1$^-$ &  1.74E+01 & 3.43E-02 &            \\
9.811 & 3.17  &  1$^-$ &  3.10E+01 & 7.99E-02 &            \\
9.982 & 3.341 &  1$^-$ &  6.29E+01 & 2.51E-02 &            \\
10.044& 3.403 &  1$^-$ &  7.38E+02 & 1.40E-01 &   2.30E-01 \\
10.099& 3.458 &  1$^-$ &  4.01E+02 & 2.36E-01 &            \\
10.142& 3.501 &  1$^-$ &  1.47E+02 & 6.71E-01 &            \\
10.173& 3.532 &  1$^-$ &  1.10E+03 & 1.14E-02 &            \\
10.186& 3.545 &  1$^-$ &  1.08E+03 & 1.71E-01 &   3.00E-01 \\
10.267& 3.626 &  1$^-$ &  9.00E+02 & 5.56E-03 &            \\
10.499& 3.858 &  1$^-$ &  2.50E+03 & 6.01E-02 &            \\
10.635& 3.994 &  1$^-$ &  4.10E+03 & 1.42E-01 &            \\
10.65 & 4.009 &  1$^-$ &  1.50E+03 & 3.44E-01 &   3.20E-01 \\
10.683& 4.042 &  1$^-$ &  6.20E+03 & 2.36E-03 &            \\
10.701& 4.06  &  1$^-$ &  5.07E+02 & 3.57E-02 &            \\
10.831& 4.19  &  1$^-$ &  5.64E+02 & 4.84E-03 &            \\
      &       &        &           &          &            \\
6.73  & 0.089 &  2$^+$ & 8.01E-03 & 4.03E-71 &            \\
6.867 & 0.226 &  2$^+$ & 8.86E-03 & 1.13E-38 &            \\
7.178 & 0.537 &  2$^+$ & 1.11E-02 & 1.55E-19 &            \\
7.423 & 0.782 &  2$^+$ & 1.31E-02 & 1.58E-13 &            \\
7.97  & 1.329 &  2$^+$ & 1.87E-02 & 7.62E-07 &            \\
8.555 & 1.914 &  2$^+$ & 2.71E-02 & 2.79E-03 &            \\
8.909 & 2.268 &  2$^+$ & 4.00E-02 & 2.99E-02 &   3.00E-02 \\
9.144 & 2.503 &  2$^+$ & 2.39E-01 & 3.20E-02 &            \\
9.356 & 2.715 &  2$^+$ & 2.69E+00 & 4.58E-02 &   5.00E-02 \\
9.373 & 2.732 &  2$^+$ & 3.48E+00 & 5.03E-02 &            \\
9.439 & 2.798 &  2$^+$ & 1.01E+01 & 2.06E-01 &            \\
9.448 & 2.807 &  2$^+$ & 5.84E+00 & 1.04E-02 &            \\
9.464 & 2.823 &  2$^+$ & 6.36E+00 & 9.70E-03 &            \\
9.502 & 2.861 &  2$^+$ & 8.74E+00 & 1.28E-02 &            \\
9.595 & 2.954 &  2$^+$ & 1.92E+01 & 7.24E-02 &            \\
9.878 & 3.237 &  2$^+$ & 9.83E+01 & 5.97E-02 &            \\
9.956 & 3.315 &  2$^+$ & 1.10E+02 & 9.00E-02 &            \\
9.9951& 3.3541&  2$^+$ & 5.02E+01 & 3.29E-01 &            \\
10.095& 3.454 &  2$^+$ & 1.79E+02 & 2.43E-03 &            \\
10.217& 3.576 &  2$^+$ & 3.10E+02 & 5.02E-02 &   5.00E-02 \\
10.319& 3.678 &  2$^+$ & 3.65E+02 & 3.18E-02 &            \\
10.328& 3.687 &  2$^+$ & 3.68E+02 & 1.11E-02 &            \\
10.346& 3.705 &  2$^+$ & 3.86E+02 & 1.16E-02 &            \\
10.439& 3.798 &  2$^+$ & 5.23E+02 & 3.85E-02 &   5.00E-01 \\
10.588& 3.947 &  2$^+$ & 6.71E+02 & 1.68E-02 &            \\
10.593& 3.952 &  2$^+$ & 6.91E+02 & 2.64E-02 &            \\
10.7  & 4.059 &  2$^+$ & 2.80E+02 & 4.08E-01 &            \\
10.789& 4.148 &  2$^+$ & 1.83E+03 & 6.08E-04 &            \\
10.851& 4.21  &  2$^+$ & 1.07E+03 & 2.49E-02 &            \\
      &       &        &          &          &            \\
6.836 & 0.195 &  3$^-$ & 2.74E-03 & 4.16E-41 &            \\
7.258 & 0.617 &  3$^-$ & 1.17E-02 & 2.17E-15 &            \\
7.627 & 0.986 &  3$^-$ & 1.50E-02 & 1.52E-08 &            \\
7.749 & 1.108 &  3$^-$ & 1.62E-02 & 4.38E-07 &            \\
8.352 & 1.711 &  3$^-$ & 2.67E-02 & 1.92E-02 &            \\
8.472 & 1.831 &  3$^-$ & 2.85E-02 & 4.96E-02 &            \\
8.672 & 2.031 &  3$^-$ & 2.90E-02 & 3.89E-03 &            \\
8.806 & 2.165 &  3$^-$ & 3.12E-02 & 2.73E-03 &            \\
9.066 & 2.425 &  3$^-$ & 1.98E-01 & 4.07E-03 &            \\
9.132 & 2.491 &  3$^-$ & 1.38E-01 & 4.91E-02 &            \\
9.192 & 2.551 &  3$^-$ & 5.29E-01 & 2.00E-01 &            \\
9.24  & 2.599 &  3$^-$ & 4.39E-01 & 1.36E-01 &            \\
9.258 & 2.617 &  3$^-$ & 5.13E-01 & 1.33E-01 &            \\
9.509 & 2.868 &  3$^-$ & 4.23E+00 & 2.10E-01 &            \\
9.737 & 3.096 &  3$^-$ & 4.00E+02 & 1.48E-03 &            \\
9.764 & 3.123 &  3$^-$ & 2.87E+01 & 2.66E-01 &            \\
9.982 & 3.341 &  3$^-$ & 5.90E+01 & 9.21E-02 &            \\
10.05 & 3.409 &  3$^-$ & 3.97E+02 & 8.58E-02 &            \\
10.167& 3.526 &  3$^-$ & 8.70E+02 & 8.94E-03 &            \\
10.255& 3.614 &  3$^-$ & 1.18E+03 & 1.18E-02 &            \\
10.28 & 3.639 &  3$^-$ & 1.25E+03 & 9.13E-05 &            \\
10.308& 3.667 &  3$^-$ & 5.40E+02 & 2.81E-02 &            \\
10.42 & 3.779 &  3$^-$ & 4.70E+03 & 4.31E-02 &            \\
10.488& 3.847 &  3$^-$ & 2.04E+02 & 4.92E-01 &   9.00E-01 \\
10.539& 3.898 &  3$^-$ & 2.59E+03 & 1.27E-01 &            \\
10.596& 3.955 &  3$^-$ & 4.57E+03 & 1.98E-01 &   3.00E-01 \\
10.674& 4.033 &  3$^-$ & 1.75E+02 & 9.08E-03 &            \\
10.853& 4.212 &  3$^-$ & 2.88E+02 & 5.07E-01 &   1.10E-01 \\
\end{tabular}
\end{table}
\newpage
\begin{table}
\caption{\label{T5} Resonance parameters for the
$^{36}$Ar($\alpha,\gamma$)$^{40}$Ca reaction.}
\begin{tabular}{rcclll}
E$_x$ & E$_r^{cm}$ & J$^{\pi}$ & $\Gamma_{tot}$ &
$\omega\gamma_{calc}$ & $\omega\gamma_{exp}$ \\
\hline
[MeV] & [MeV] &  & [eV] & [eV] & [eV]\\
\hline
7.3   & 0.259 &  0$^+$ & 3.80E-03 & 9.85E-40 &            \\
7.701 & 0.66  &  0$^+$ & 4.88E-02 & 2.55E-18 &            \\
7.814 & 0.773 &  0$^+$ & 5.24E-02 & 5.26E-16 &            \\
8.018 & 0.977 &  0$^+$ & 5.96E-02 & 2.24E-12 &            \\
8.27  & 1.229 &  0$^+$ & 6.96E-02 & 3.18E-09 &            \\
8.439 & 1.398 &  0$^+$ & 7.70E-02 & 6.19E-08 &            \\
8.484 & 1.443 &  0$^+$ & 1.90E-02 & 1.61E-07 &            \\
8.938 & 1.897 &  0$^+$ & 1.03E-01 & 5.53E-05 &            \\
9.304 & 2.263 &  0$^+$ & 2.15E-01 & 1.61E-03 &            \\
10.42 & 3.379 &  0$^+$ & 4.55E+01 & 1.87E-01 &            \\
12.27 & 5.229 &  0$^+$ & 4.10E+03 & 1.07E-01 &            \\
      &       &        &          &          &            \\
7.113 & 0.072 &  1$^-$ & 8.20E-03 & 1.13E-90 &            \\
8.113 & 1.072 &  1$^-$ & 1.19E-02 & 2.20E-10 &            \\
8.271 & 1.23  &  1$^-$ & 6.97E-02 & 6.07E-09 &            \\
8.323 & 1.282 &  1$^-$ & 8.20E-03 & 2.04E-08 &            \\
8.358 & 1.317 &  1$^-$ & 4.40E-03 & 4.43E-08 &            \\
8.665 & 1.624 &  1$^-$ & 8.80E-02 & 6.27E-06 &            \\
8.994 & 1.953 &  1$^-$ & 7.01E-02 & 2.33E-03 &            \\
9.432 & 2.391 &  1$^-$ & 2.30E+02 & 6.58E-05 &            \\
9.538 & 2.497 &  1$^-$ & 4.00E+02 & 5.12E-06 &            \\
9.604 & 2.563 &  1$^-$ & 1.90E+02 & 4.18E-04 &            \\
10.199& 3.158 &  1$^-$ & 2.70E+02 & 8.04E-04 &            \\
10.267& 3.226 &  1$^-$ & 1.77E+02 & 7.06E-04 &            \\
10.278& 3.237 &  1$^-$ & 1.03E+03 & 8.70E-04 &            \\
10.421& 3.38  &  1$^-$ & 5.94E+02 & 4.70E-03 &            \\
10.623& 3.582 &  1$^-$ & 5.00E+03 & 6.78E-04 &            \\
10.67 & 3.629 &  1$^-$ & 5.70E+03 & 1.48E-04 &            \\
10.91 & 3.869 &  1$^-$ & 2.30E+03 & 1.42E-02 &            \\
10.951& 3.91  &  1$^-$ & 1.00E+04 & 3.47E-02 &            \\
11.022& 3.981 &  1$^-$ & 2.71E+02 & 2.92E-01 &            \\
12.099& 5.058 &  1$^-$ & 1.09E+03 & 1.50E-01 &   1.00E-01 \\
12.944& 5.903 &  1$^-$ & 3.78E+03 & 1.12E+00 &   3.40E+00 \\
13.245& 6.204 &  1$^-$ & 9.30E+03 & 1.15E+00 &   9.70E+00 \\
13.442& 6.401 &  1$^-$ & 5.63E+03 & 1.52E+00 &   3.40E+00 \\
13.784& 6.743 &  1$^-$ & 7.07E+03 & 1.83E+00 &   3.70E+00 \\
13.956& 6.915 &  1$^-$ & 2.40E+04 & 1.47E+01 &   1.46E+01 \\
14.08 & 7.039 &  1$^-$ & 1.43E+04 & 1.69E+01 &   1.44E+01 \\
14.42 & 7.379 &  1$^-$ & 1.71E+04 & 2.83E+00 &   4.70E+00 \\
14.51 & 7.469 &  1$^-$ & 1.79E+04 & 2.93E+00 &   4.50E+00 \\
14.87 & 7.829 &  1$^-$ & 2.09E+04 & 3.36E+00 &   6.30E+00 \\
      &       &        &          &          &            \\
7.277 & 0.236 &  2$^+$ & 9.30E-03 & 3.36E-43 &            \\
7.446 & 0.405 &  2$^+$ & 5.90E-02 & 8.90E-28 &            \\
7.676 & 0.635 &  2$^+$ & 2.00E-03 & 1.85E-19 &            \\
7.872 & 0.831 &  2$^+$ & 2.25E-01 & 6.39E-15 &            \\
7.976 & 0.935 &  2$^+$ & 1.00E-02 & 7.85E-13 &            \\
8.091 & 1.05  &  2$^+$ & 1.48E-01 & 3.77E-11 &            \\
8.338 & 1.297 &  2$^+$ & 7.25E-02 & 2.92E-08 &            \\
8.578 & 1.537 &  2$^+$ & 9.30E-02 & 3.69E-06 &            \\
8.748 & 1.707 &  2$^+$ & 6.50E-02 & 5.58E-05 &            \\
8.934 & 1.893 &  2$^+$ & 2.58E-01 & 2.95E-04 &            \\
8.981 & 1.94  &  2$^+$ & 5.40E-02 & 5.72E-04 &            \\
9.227 & 2.186 &  2$^+$ & 4.97E+00 & 3.94E-04 &            \\
9.388 & 2.347 &  2$^+$ & 2.11E+01 & 3.86E-04 &            \\
10.54 & 3.499 &  2$^+$ & 5.07E+02 & 1.96E-02 &            \\
10.691& 3.65  &  2$^+$ & 1.40E+02 & 2.53E-02 &            \\
10.78 & 3.739 &  2$^+$ & 1.80E+02 & 1.79E-02 &            \\
10.862& 3.821 &  2$^+$ & 4.53E+01 & 1.21E-01 &            \\
11.042& 4.001 &  2$^+$ & 5.00E+02 & 1.45E-02 &            \\
11.117& 4.076 &  2$^+$ & 3.18E+01 & 3.68E+00 &            \\
11.266& 4.225 &  2$^+$ & 3.40E+02 & 1.20E+00 &            \\
11.324& 4.283 &  2$^+$ & 5.76E+02 & 2.77E-01 &            \\
11.369& 4.328 &  2$^+$ & 2.47E+02 & 8.29E-01 &            \\
11.459& 4.418 &  2$^+$ & 1.31E+03 & 5.38E-01 &            \\
      &       &        &          &          &            \\
7.239 & 0.198 &  3$^-$ & 4.66E-03 & 2.10E-49 &            \\
7.623 & 0.582 &  3$^-$ & 4.07E-03 & 9.94E-22 &            \\
7.769 & 0.728 &  3$^-$ & 2.72E-03 & 2.25E-16 &            \\
8.764 & 1.723 &  3$^-$ & 9.65E-02 & 2.60E-05 &            \\
9.091 & 2.05  &  3$^-$ & 1.25E-01 & 9.54E-04 &            \\
9.362 & 2.321 &  3$^-$ & 1.27E+00 & 2.49E-03 &            \\
9.453 & 2.412 &  3$^-$ & 9.00E+01 & 5.87E-05 &            \\
10.13 & 3.089 &  3$^-$ & 7.32E+01 & 2.17E-02 &            \\
10.262& 3.221 &  3$^-$ & 3.74E+02 & 8.56E-04 &            \\
10.361& 3.32  &  3$^-$ & 3.90E+03 & 7.92E-05 &            \\
10.443& 3.402 &  3$^-$ & 4.40E+02 & 1.09E-02 &            \\
10.776& 3.735 &  3$^-$ & 1.20E+03 & 1.07E-01 &            \\
11.011& 3.97  &  3$^-$ & 2.76E+02 & 7.93E-02 &            \\
11.249& 4.208 &  3$^-$ & 8.47E+02 & 8.40E-02 &            \\
\end{tabular}
\end{table}
\newpage
\begin{table}
\caption{\label{T6} Resonance parameters for the
$^{40}$Ca($\alpha,\gamma$)$^{44}$Ti reaction.}
\begin{tabular}{rcclll}
E$_x$ & E$_r^{cm}$ & J$^{\pi}$ & $\Gamma_{tot}$ &
$\omega\gamma_{calc}$ & $\omega\gamma_{exp}$ \\
\hline
[MeV] & [MeV] &  & [eV] & [eV] & [eV]\\
\hline
8.954 & 3.827 &  1$^-$ & 5.38E-01 & 2.34E-01 &   2.20E-01  \\
      &       &        &          &          &             \\
6.22  & 1.093 &  2$^+$ & 7.17E-02 & 2.81E-13 &             \\
7.634 & 2.507 &  2$^+$ & 2.02E-01 & 1.32E-02 &   1.30E-02  \\
8.067 & 2.94  &  2$^+$ & 2.68E-01 & 2.27E-02 &   2.20E-02  \\
8.318 & 3.191 &  2$^+$ & 3.37E-01 & 1.38E-01 &   1.20E-01  \\
8.385 & 3.258 &  2$^+$ & 4.76E-01 & 5.26E-01 &   5.20E-01  \\
8.449 & 3.322 &  2$^+$ & 3.96E-01 & 2.69E-01 &   2.80E-01  \\
8.511 & 3.384 &  2$^+$ & 4.11E-01 & 2.82E-01 &   2.20E-01  \\
8.534 & 3.407 &  2$^+$ & 4.50E-01 & 3.93E-01 &   3.30E-01  \\
8.565 & 3.438 &  2$^+$ & 3.78E-01 & 1.06E-01 &   1.10E-01  \\
8.627 & 3.5   &  2$^+$ & 3.84E-01 & 7.87E-02 &   8.00E-02  \\
8.639 & 3.512 &  2$^+$ & 4.40E-01 & 2.94E-01 &   2.30E-01  \\
8.756 & 3.629 &  2$^+$ & 4.73E-01 & 3.22E-01 &   3.30E-01  \\
8.946 & 3.819 &  2$^+$ & 4.70E-01 & 1.37E-01 &   1.10E-01  \\
8.987 & 3.86  &  2$^+$ & 5.38E-01 & 3.64E-01 &   3.00E-01  \\
9.215 & 4.088 &  2$^+$ & 6.23E-01 & 4.43E-01 &   5.00E-01  \\
9.227 & 4.1   &  2$^+$ & 7.04E+00 & 5.41E+00 &   5.80E+00  \\
9.239 & 4.112 &  2$^+$ & 2.90E+00 & 2.12E+00 &   2.00E+00  \\
9.361 & 4.234 &  2$^+$ & 1.18E+00 & 1.31E+00 &   1.20E+00  \\
10.386& 5.259 &  2$^+$ & 4.99E+02 & 4.95E+00 &   5.00E+00  \\
      &       &        &          &          &             \\
8.534 & 3.407 &  3$^-$ & 4.09E-01 & 3.61E-01 &   3.30E-01  \\
8.96  & 3.833 &  3$^-$ & 5.25E-01 & 4.78E-01 &   4.00E-01  \\
9.361 & 4.234 &  3$^-$ & 9.43E-01 & 1.33E+00 &   1.20E+00  \\
10.386& 5.259 &  3$^-$ & 1.15E+03 & 4.86E+00 &   5.00E+00  \\
      &       &        &          &          &             \\
5.21  & 0.083 &  4$^+$ & 1.32E-03 & 2.68E-61 &             \\
5.305 & 0.178 &  4$^+$ & 1.32E-03 & 2.68E-61 &             \\
8.992 & 3.865 &  4$^+$ & 5.32E-01 & 6.11E-01 &   6.00E-01  \\
9.427 & 4.3   &  4$^+$ & 9.08E-01 & 8.95E-01 &   9.00E-01  \\
9.713 & 4.586 &  4$^+$ & 8.86E+00 & 2.33E+00 &   2.50E+00  \\
\end{tabular}
\end{table}
\newpage
\begin{table}
\caption{\label{T7} Reaction rate for $^{20}$Ne($\alpha,\gamma$)$^{24}$Mg as
a function of temperature. Listed are the rates calculated on the basis of
measured resonances,
and of the predicted resonances using the parameters given in table \ref{T1}.
For comparison
also shown is the rate predicted with the statistical model code NON-SMOKER. }
\begin{tabular}{rcclll}
Temperature& N$_A<\sigma v>_{exp}$ & N$_A<\sigma v>_{emp}$ & N$_A<\sigma v>_{HF}$ \\
\hline
[GK] & [cm$^3$s$^{-1}$mole$^{-1}$] & [cm$^3$s$^{-1}$mole$^{-1}$] & [cm$^3$s$^{-1}$mole$^{-1}$] \\
\hline
   .1000  &       1.40E-38  &    9.13E-28  &      5.77E-25   \\
   .1500  &       1.94E-25  &    2.65E-24  &      2.55E-19   \\
   .2000  &       6.38E-19  &    5.89E-19  &      6.48E-16   \\
   .3000  &       1.75E-12  &    1.67E-12  &      1.10E-11   \\
   .4000  &       2.56E-09  &    2.53E-09  &      4.58E-09   \\
   .5000  &       1.90E-07  &    2.01E-07  &      3.02E-07   \\
   .6000  &       3.25E-06  &    3.88E-06  &      6.77E-06   \\
   .7000  &       2.50E-05  &    3.50E-05  &      7.46E-05   \\
   .8000  &       1.26E-04  &    2.03E-04  &      5.01E-04   \\
   .9000  &       5.22E-04  &    9.04E-04  &      2.35E-03   \\
  1.0000  &       1.94E-03  &    3.34E-03  &      8.49E-03   \\
  1.5000  &       2.47E-01  &    3.36E-01  &      5.35E-01   \\
  2.0000  &       3.43E+00  &    4.43E+00  &      5.33E+00   \\
  3.0000  &       4.60E+01  &    6.28E+01  &      6.60E+01   \\
  4.0000  &       1.58E+02  &    2.39E+02  &      2.58E+02   \\
  5.0000  &       3.16E+02  &    5.35E+02  &      6.06E+02   \\
  6.0000  &       4.83E+02  &    9.13E+02  &      1.09E+03   \\
  7.0000  &       6.37E+02  &    1.33E+03  &      1.66E+03   \\
  8.0000  &       7.67E+02  &    1.76E+03  &      2.28E+03   \\
  9.0000  &       8.70E+02  &    2.16E+03  &      2.92E+03   \\
 10.0000  &       9.48E+02  &    2.54E+03  &      3.55E+03   \\
\end{tabular}
\end{table}
\newpage
\begin{table}
\caption{\label{T8} Reaction rate for $^{24}$Mg($\alpha,\gamma$)$^{28}$Si as
a function of temperature. Listed are the rates calculated on the basis of measured resonances,
and of the predicted resonances using the parameters given in table \ref{T2}. For comparison
also shown is the rate predicted with the statistical model code NON-SMOKER. }
\begin{tabular}{rcclll}
Temperature& N$_A<\sigma v>_{exp}$ & N$_A<\sigma v>_{emp}$ & N$_A<\sigma v>_{HF}$ \\
\hline
[GK] & [cm$^3$s$^{-1}$mole$^{-1}$] & [cm$^3$s$^{-1}$mole$^{-1}$] & [cm$^3$s$^{-1}$mole$^{-1}$] \\
\hline
   .1000  &      0.00E+00   &    8.02E-32   &     7.85E-30   \\
   .1500  &      4.52E-37   &    1.61E-24   &     2.08E-23   \\
   .2000  &      1.90E-27   &    3.19E-20   &     1.60E-19   \\
   .3000  &      6.85E-18   &    1.53E-15   &     1.13E-14   \\
   .4000  &      3.82E-13   &    4.11E-12   &     1.22E-11   \\
   .5000  &      2.68E-10   &    1.19E-09   &     1.64E-09   \\
   .6000  &      2.16E-08   &    8.62E-08   &     6.44E-08   \\
   .7000  &      5.17E-07   &    2.19E-06   &     1.13E-06   \\
   .8000  &      5.87E-06   &    2.60E-05   &     1.13E-05   \\
   .9000  &      4.09E-05   &    1.80E-04   &     7.45E-05   \\
  1.0000  &      2.02E-04   &    8.54E-04   &     3.58E-04   \\
  1.5000  &      3.24E-02   &    9.50E-02   &     5.73E-02   \\
  2.0000  &      4.80E-01   &    1.06E+00   &     9.14E-01   \\
  3.0000  &      8.42E+00   &    1.34E+01   &     1.71E+01   \\
  4.0000  &      3.98E+01   &    5.49E+01   &     7.86E+01   \\
  5.0000  &      1.05E+02   &    1.43E+02   &     2.00E+02   \\
  6.0000  &      1.99E+02   &    2.86E+02   &     3.76E+02   \\
  7.0000  &      3.11E+02   &    4.79E+02   &     5.95E+02   \\
  8.0000  &      4.27E+02   &    7.09E+02   &     8.47E+02   \\
  9.0000  &      5.40E+02   &    9.59E+02   &     1.12E+03   \\
 10.0000  &      6.42E+02   &    1.21E+03   &     1.41E+03   \\
\end{tabular}
\end{table}
\newpage
\begin{table}
\caption{\label{T9} Reaction rate for $^{28}$Si($\alpha,\gamma$)$^{32}$S as
a function of temperature. Listed are the rates calculated on the basis of measured resonances,
and of the predicted resonances using the parameters given in table \ref{T3}. For comparison
also shown is the rate predicted with the statistical model code NON-SMOKER. }
\begin{tabular}{rcclll}
Temperature& N$_A<\sigma v>_{exp}$ & N$_A<\sigma v>_{emp}$ & N$_A<\sigma v>_{HF}$ \\
\hline
[GK] & [cm$^3$s$^{-1}$mole$^{-1}$] & [cm$^3$s$^{-1}$mole$^{-1}$] & [cm$^3$s$^{-1}$mole$^{-1}$] \\
\hline
   .1000   &     0.00E+00   &   1.42E-36   &    2.20E-34    \\
   .1500   &     0.00E+00   &   1.44E-28   &    3.14E-27    \\
   .2000   &     4.90E-36   &   1.67E-24   &    6.94E-23    \\
   .3000   &     2.54E-23   &   7.37E-20   &    1.83E-17    \\
   .4000   &     5.09E-17   &   1.12E-16   &    4.64E-14    \\
   .5000   &     2.86E-13   &   3.11E-13   &    1.17E-11    \\
   .6000   &     8.67E-11   &   9.09E-11   &    7.45E-10    \\
   .7000   &     4.98E-09   &   5.16E-09   &    1.91E-08    \\
   .8000   &     1.02E-07   &   1.05E-07   &    2.60E-07    \\
   .9000   &     1.06E-06   &   1.08E-06   &    2.20E-06    \\
  1.0000   &     6.80E-06   &   6.97E-06   &    1.30E-05    \\
  1.5000   &     1.90E-03   &   1.98E-03   &    4.13E-03    \\
  2.0000   &     4.00E-02   &   4.23E-02   &    9.84E-02    \\
  3.0000   &     1.34E+00   &   1.45E+00   &    3.04E+00    \\
  4.0000   &     9.39E+00   &   1.16E+01   &    1.90E+01    \\
  5.0000   &     3.15E+01   &   4.65E+01   &    5.97E+01    \\
  6.0000   &     7.07E+01   &   1.22E+02   &    1.31E+02    \\
  7.0000   &     1.24E+02   &   2.45E+02   &    2.35E+02    \\
  8.0000   &     1.87E+02   &   4.11E+02   &    3.67E+02    \\
  9.0000   &     2.54E+02   &   6.07E+02   &    5.24E+02    \\
 10.0000   &     3.20E+02   &   8.21E+02   &    7.01E+02    \\
\end{tabular}
\end{table}
\newpage
\begin{table}
\caption{\label{T10} Reaction rate for $^{32}$S($\alpha,\gamma$)$^{36}$Ar as
a function of temperature. Listed are the rates calculated on the basis of measured resonances,
and of the predicted resonances using the parameters given in table \ref{T4}. For comparison
also shown is the rate predicted with the statistical model code NON-SMOKER. }
\begin{tabular}{rcclll}
Temperature& N$_A<\sigma v>_{exp}$ & N$_A<\sigma v>_{emp}$ & N$_A<\sigma v>_{HF}$ \\
\hline
[GK] & [cm$^3$s$^{-1}$mole$^{-1}$] & [cm$^3$s$^{-1}$mole$^{-1}$] & [cm$^3$s$^{-1}$mole$^{-1}$] \\
\hline
   .1000  &       0.00E+00   &   5.78E-40   &     3.70E-39    \\
   .1500  &       0.00E+00   &   1.71E-30   &     2.64E-31    \\
   .2000  &       0.00E+00   &   1.63E-25   &     1.60E-26    \\
   .3000  &       4.55E-35   &   8.78E-20   &     1.50E-20    \\
   .4000  &       1.04E-25   &   9.70E-16   &     8.52E-17    \\
   .5000  &       4.12E-20   &   3.21E-13   &     3.86E-14    \\
   .6000  &       2.18E-16   &   2.03E-11   &     3.93E-12    \\
   .7000  &       9.95E-14   &   7.43E-10   &     1.51E-10    \\
   .8000  &       9.86E-12   &   1.72E-08   &     2.91E-09    \\
   .9000  &       3.53E-10   &   2.25E-07   &     3.39E-08    \\
  1.0000  &       6.19E-09   &   1.81E-06   &     2.70E-07    \\
  1.5000  &       3.30E-05   &   9.43E-04   &     2.41E-04    \\
  2.0000  &       2.32E-03   &   2.22E-02   &     9.64E-03    \\
  3.0000  &       1.53E-01   &   6.44E-01   &     4.56E-01    \\
  4.0000  &       1.23E+00   &   4.02E+00   &     3.42E+00    \\
  5.0000  &       4.34E+00   &   1.26E+01   &     1.23E+01    \\
  6.0000  &       1.01E+01   &   2.71E+01   &     3.07E+01    \\
  7.0000  &       1.84E+01   &   4.65E+01   &     6.25E+01    \\
  8.0000  &       2.88E+01   &   6.91E+01   &     1.12E+02    \\
  9.0000  &       4.03E+01   &   9.29E+01   &     1.81E+02    \\
 10.0000  &       5.22E+01   &   1.17E+02   &     2.73E+02    \\
\end{tabular}
\end{table}
\newpage
\begin{table}
\caption{\label{T11} Reaction rate for $^{36}$Ar($\alpha,\gamma$)$^{40}$Ca as
a function of temperature. Listed are the rates calculated on the basis of measured resonances,
and of the predicted resonances using the parameters given in table \ref{T5}. For comparison
also shown is the rate predicted with the statistical model code NON-SMOKER. }
\begin{tabular}{rcclll}
Temperature& N$_A<\sigma v>_{exp}$ & N$_A<\sigma v>_{emp}$ & N$_A<\sigma v>_{HF}$ \\
\hline
[GK] & [cm$^3$s$^{-1}$mole$^{-1}$] & [cm$^3$s$^{-1}$mole$^{-1}$] & [cm$^3$s$^{-1}$mole$^{-1}$] \\
\hline
   .1000   &       0.00E+00  &    2.46E-42   &    2.07E-43    \\
   .1500   &       0.00E+00  &    1.51E-34   &    6.84E-35    \\
   .2000   &       0.00E+00  &    5.27E-29   &    1.08E-29    \\
   .3000   &       0.00E+00  &    1.18E-22   &    3.41E-23    \\
   .4000   &       0.00E+00  &    1.71E-18   &    4.08E-19    \\
   .5000   &       0.00E+00  &    1.21E-15   &    3.12E-16    \\
   .6000   &       2.80E-45  &    1.55E-13   &    4.85E-14    \\
   .7000   &       2.67E-38  &    7.33E-12   &    2.52E-12    \\
   .8000   &       4.85E-33  &    1.76E-10   &    5.83E-11    \\
   .9000   &       5.93E-29  &    2.42E-09   &    7.44E-10    \\
  1.0000   &       1.10E-25  &    2.11E-08   &    6.07E-09    \\
  1.5000   &       6.67E-16  &    1.71E-05   &    4.91E-06    \\
  2.0000   &       5.07E-11  &    5.01E-04   &    1.95E-04    \\
  3.0000   &       3.72E-06  &    1.93E-02   &    1.39E-02    \\
  4.0000   &       9.77E-04  &    2.31E-01   &    2.03E-01    \\
  5.0000   &       2.70E-02  &    1.35E+00   &    1.41E+00    \\
  6.0000   &       2.40E-01  &    4.60E+00   &    6.02E+00    \\
  7.0000   &       1.12E+00  &    1.12E+01   &    1.85E+01    \\
  8.0000   &       3.49E+00  &    2.18E+01   &    4.46E+01    \\
  9.0000   &       8.30E+00  &    3.68E+01   &    9.08E+01    \\
 10.0000   &       1.64E+01  &    5.62E+01   &    1.62E+02    \\
\end{tabular}
\end{table}
\newpage
\begin{table}
\caption{\label{T12} Reaction rate for $^{40}$Ca($\alpha,\gamma$)$^{44}$Ti as
a function of temperature. Listed are the rates calculated on the basis of measured resonances,
and of the predicted resonances using the parameters given in table \ref{T6}. For comparison
also shown is the rate predicted with the statistical model code NON-SMOKER. }
\begin{tabular}{rcclll}
Temperature& N$_A<\sigma v>_{exp}$ & N$_A<\sigma v>_{emp}$ & N$_A<\sigma v>_{HF}$ \\
\hline
[GK] & [cm$^3$s$^{-1}$mole$^{-1}$] & [cm$^3$s$^{-1}$mole$^{-1}$] & [cm$^3$s$^{-1}$mole$^{-1}$] \\
\hline
   .1000   &      0.00E+00   &   0.00E+00   &     1.63E-47    \\
   .1500   &      0.00E+00   &   1.96E-44   &     2.35E-38    \\
   .2000   &      0.00E+00   &   2.00E-35   &     9.41E-33    \\
   .3000   &      1.34E-39   &   1.65E-26   &     9.48E-26    \\
   .4000   &      2.95E-29   &   4.17E-22   &     2.39E-21    \\
   .5000   &      4.38E-23   &   1.70E-19   &     3.16E-18    \\
   .6000   &      5.43E-19   &   9.40E-18   &     7.43E-16    \\
   .7000   &      4.40E-16   &   5.90E-16   &     5.71E-14    \\
   .8000   &      6.51E-14   &   6.73E-14   &     2.01E-12    \\
   .9000   &      3.13E-12   &   3.18E-12   &     3.96E-11    \\
  1.0000   &      6.87E-11   &   6.97E-11   &     5.02E-10    \\
  1.5000   &      8.25E-07   &   8.29E-07   &     2.67E-06    \\
  2.0000   &      1.33E-04   &   1.37E-04   &     3.63E-04    \\
  3.0000   &      3.19E-02   &   3.86E-02   &     8.37E-02    \\
  4.0000   &      5.29E-01   &   7.47E-01   &     1.64E+00    \\
  5.0000   &      2.85E+00   &   4.49E+00   &     1.08E+01    \\
  6.0000   &      8.74E+00   &   1.47E+01   &     3.92E+01    \\
  7.0000   &      1.93E+01   &   3.39E+01   &     1.01E+02    \\
  8.0000   &      3.47E+01   &   6.27E+01   &     2.06E+02    \\
  9.0000   &      5.42E+01   &   9.97E+01   &     3.63E+02    \\
 10.0000   &      7.69E+01   &   1.43E+02   &     5.75E+02    \\
\end{tabular}
\end{table}
\newpage


\clearpage


\begin{figure}[htp]
\begin{center}
\epsfig{file=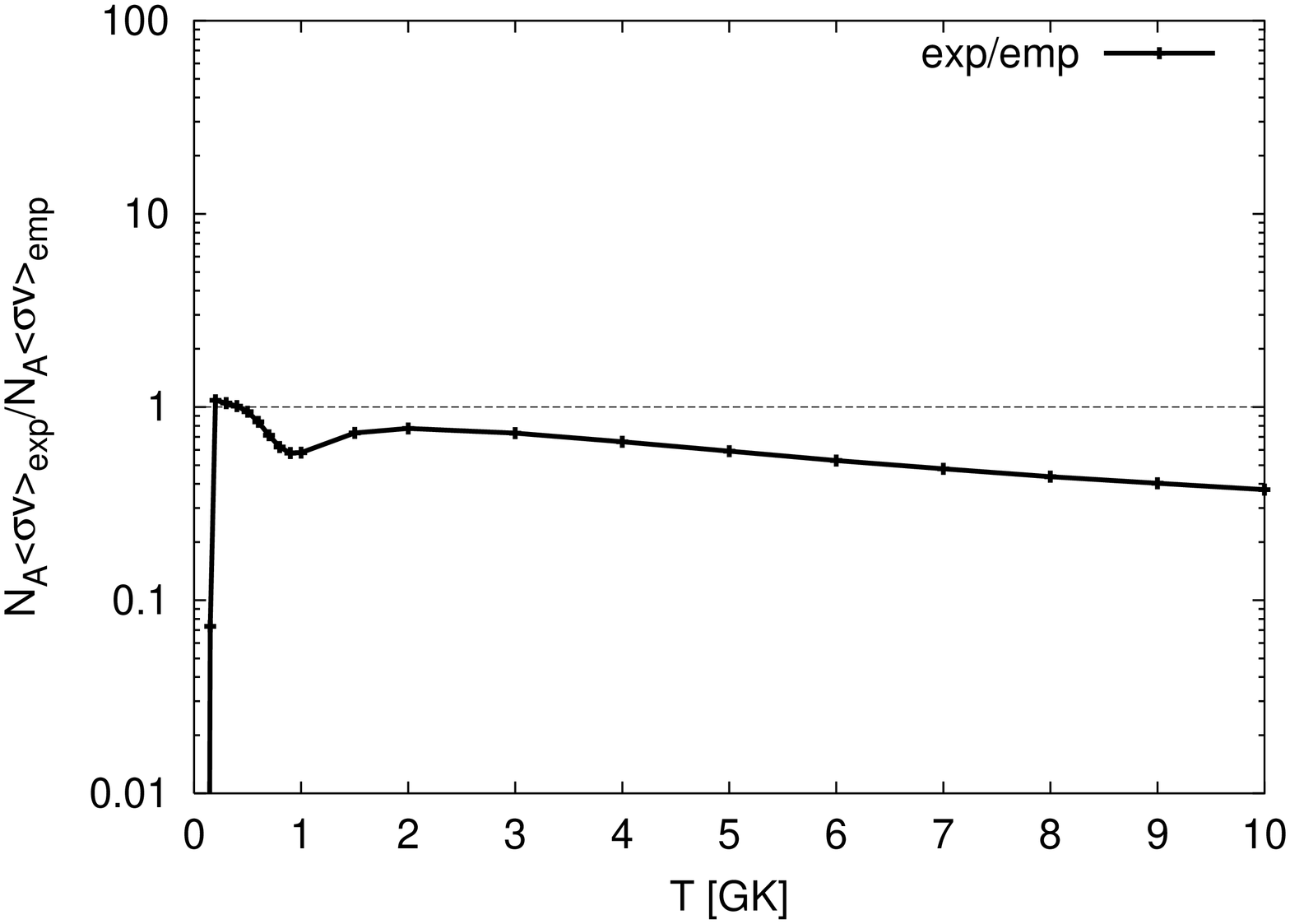,width=10cm,angle=-90}
\caption{Ratio of the 'experimental' rate (directly derived from $\alpha$
capture data) and the 'empirical' rate (derived from different sources; see
text) for $^{20}$Ne($\alpha$,$\gamma$). The temperature $T$ is given in GK.}
\label{fig:ne20}
\end{center}
\end{figure}

\begin{figure}[htp]
\begin{center}
\epsfig{file=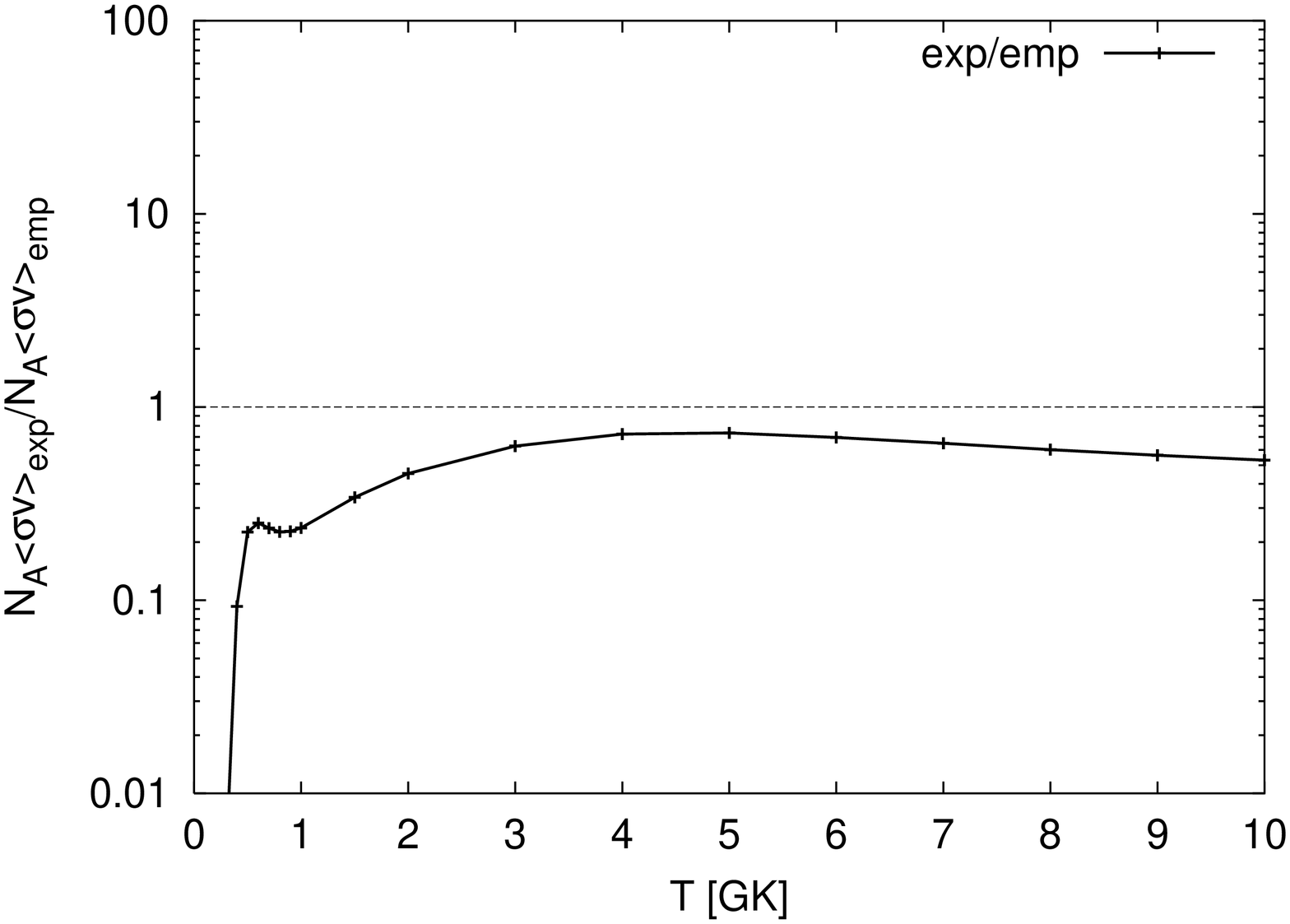,width=10cm,angle=-90}
\caption{Ratio of the 'experimental' rate (directly derived from $\alpha$
capture data) and the 'empirical' rate (derived from different sources; see
text) for $^{24}$Mg($\alpha$,$\gamma$). The temperature $T$ is given in GK.}
\label{fig:mg24}\end{center}
\end{figure}

\begin{figure}[htp]
\begin{center}
\epsfig{file=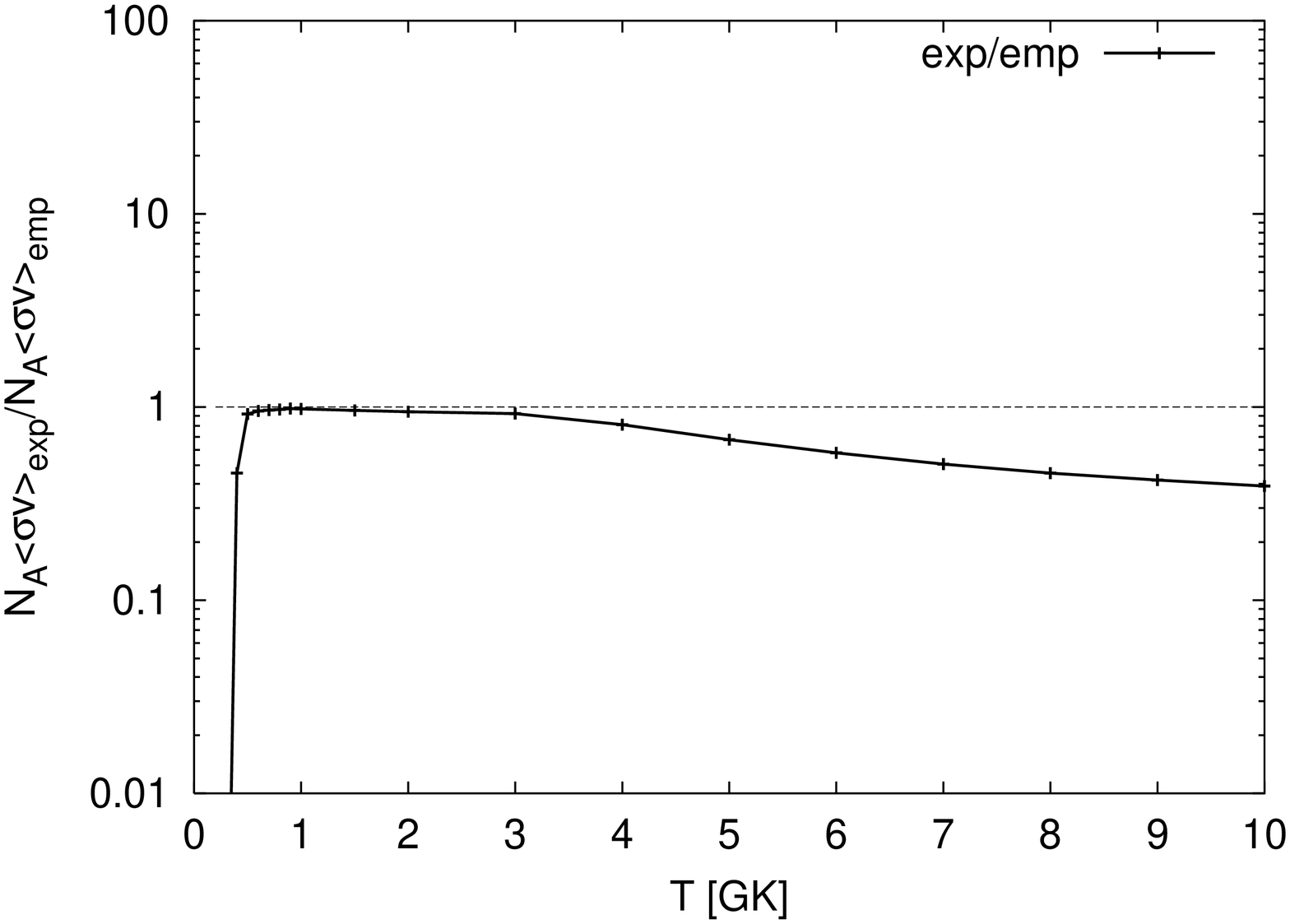,width=10cm,angle=-90}
\caption{Ratio of the 'experimental' rate (directly derived from $\alpha$
capture data) and the 'empirical' rate (derived from different sources; see
text) for $^{28}$Si($\alpha$,$\gamma$). The temperature $T$ is given in GK.}
\label{fig:si28}\end{center}
\end{figure}

\begin{figure}[htp]
\begin{center}
\epsfig{file=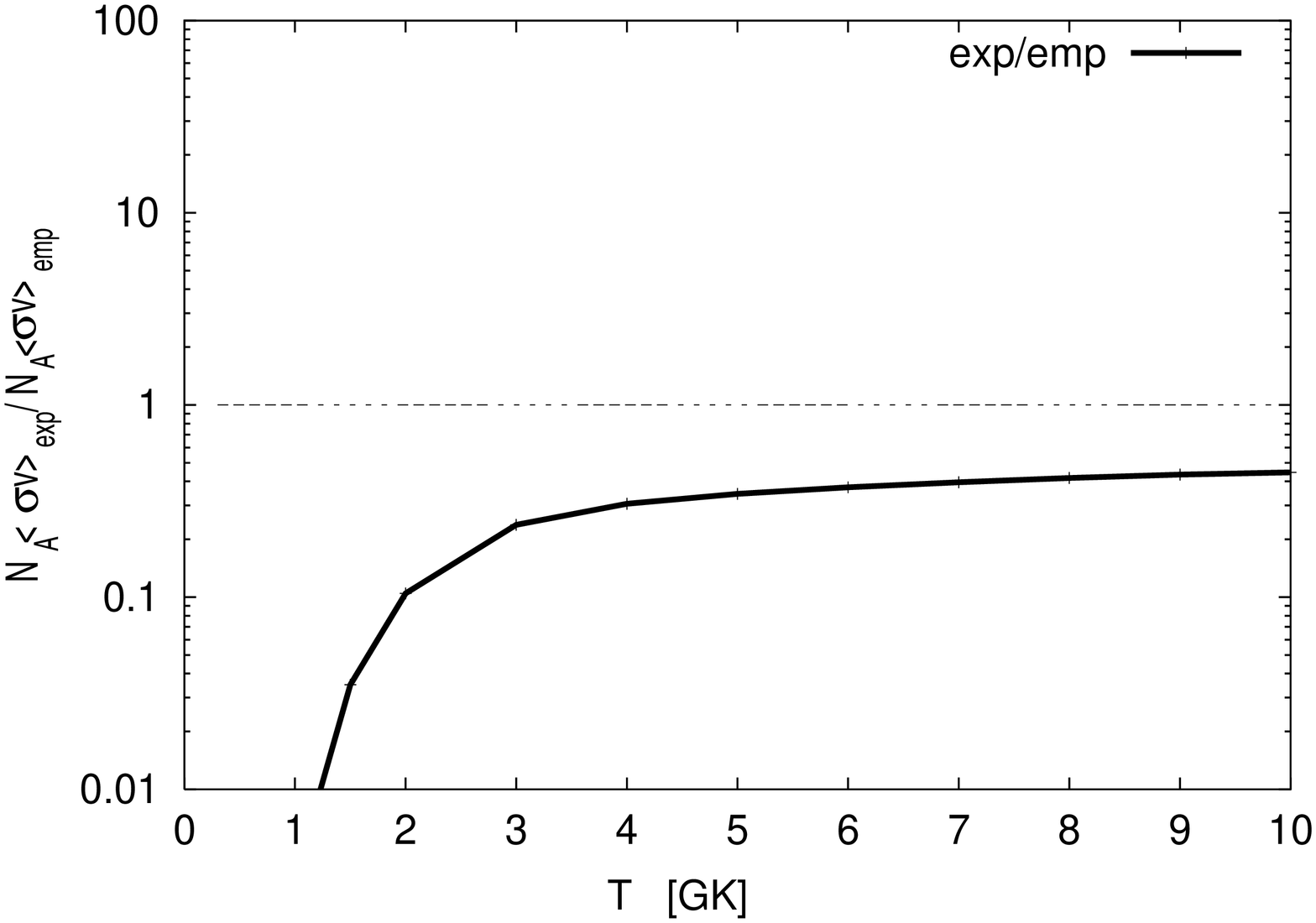,width=10cm,angle=-90}
\caption{Ratio of the 'experimental' rate (directly derived from $\alpha$
capture data) and the 'empirical' rate (derived from different sources; see
text) for $^{32}$S($\alpha$,$\gamma$). The temperature $T$ is given in GK.}
\label{fig:s32}\end{center}
\end{figure}

\begin{figure}[htp]
\begin{center}
\epsfig{file=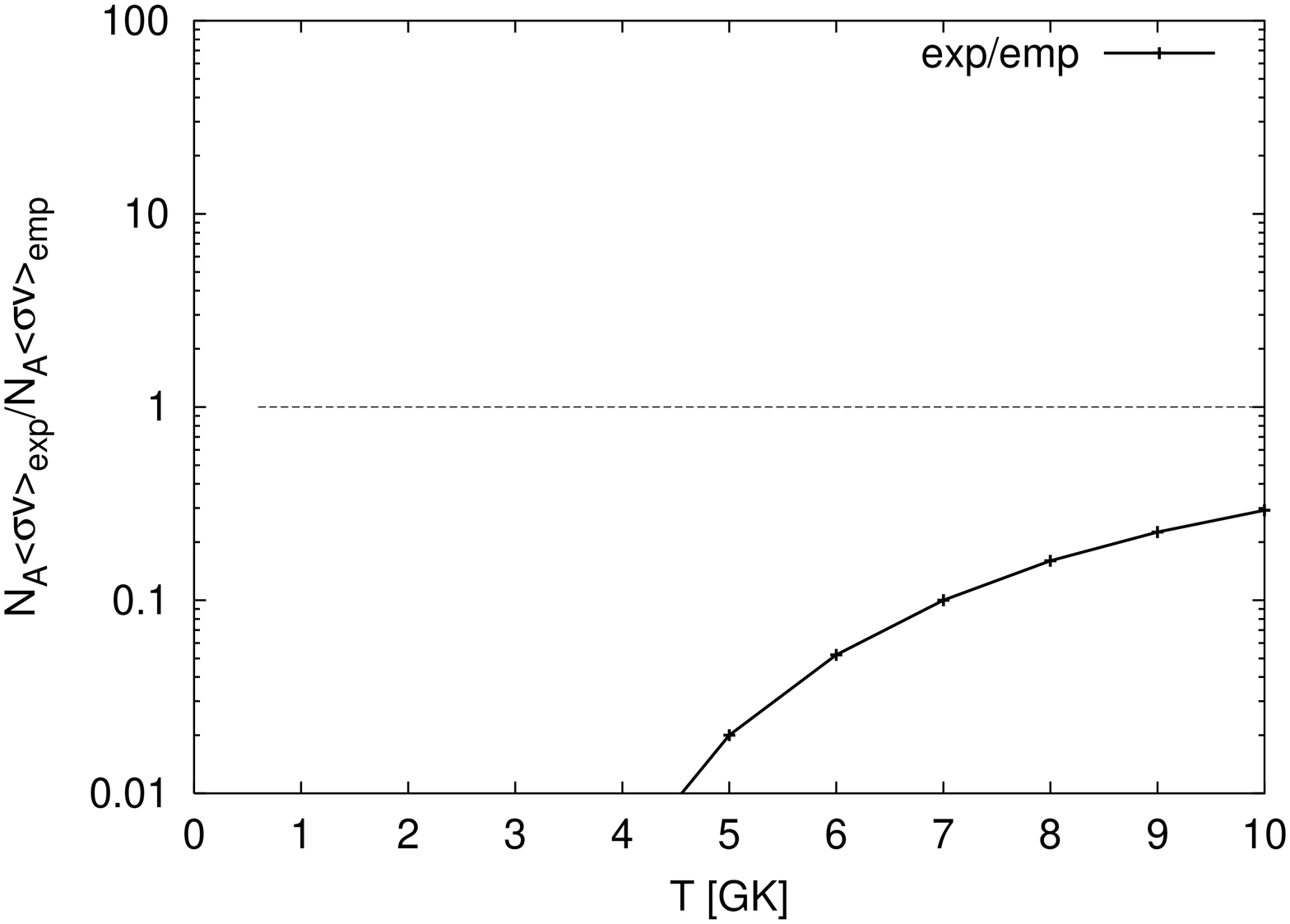,width=10cm,angle=-90}
\caption{Ratio of the 'experimental' rate (directly derived from $\alpha$
capture data) and the 'empirical' rate (derived from different sources; see
text) for $^{36}$Ar($\alpha$,$\gamma$). The temperature $T$ is given in GK.}
\label{fig:ar36}\end{center}
\end{figure}

\begin{figure}[htp]
\begin{center}
\epsfig{file=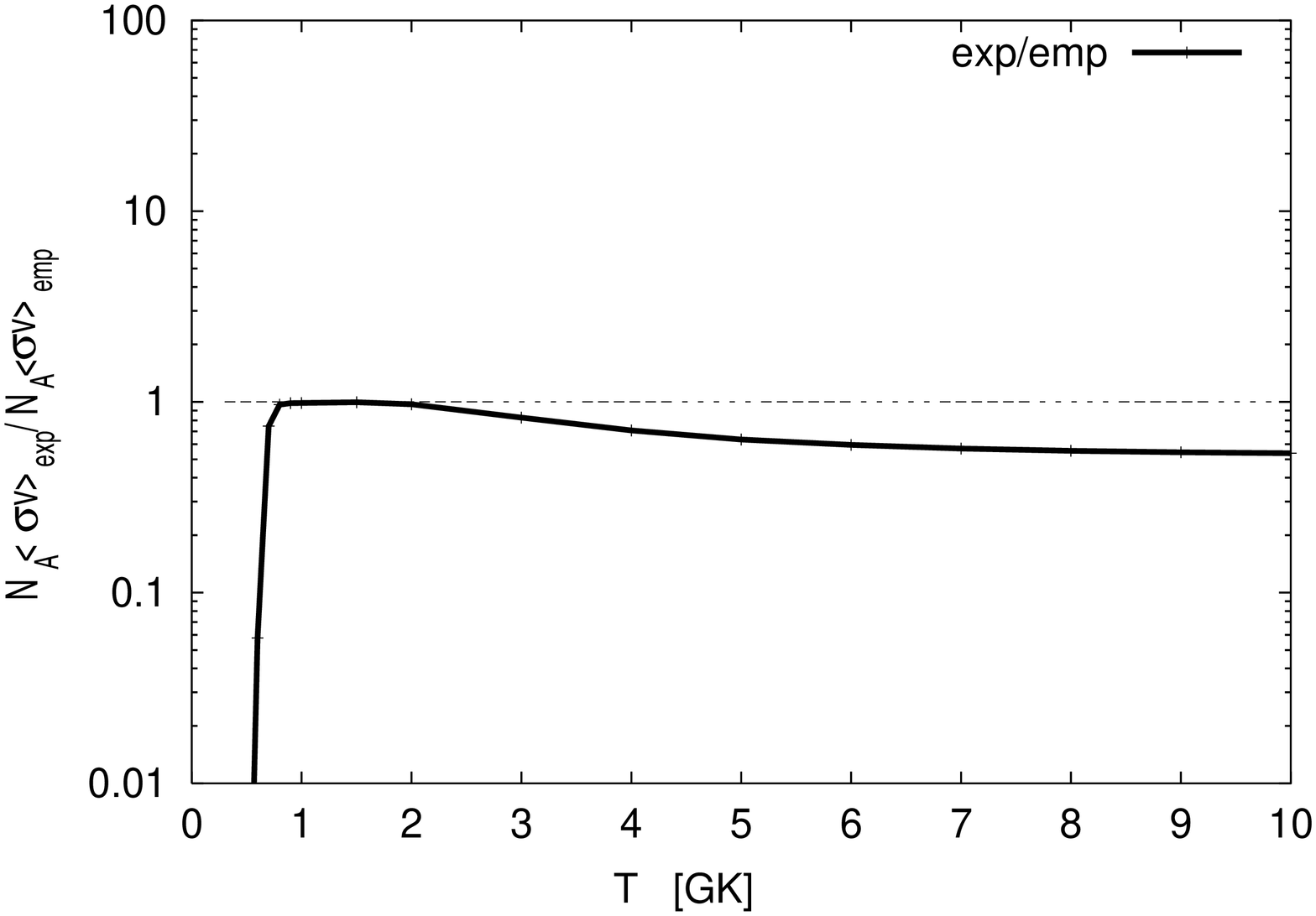,width=10cm,angle=-90}
\caption{Ratio of the 'experimental' rate (directly derived from $\alpha$
capture data) and the 'empirical' rate (derived from different sources; see
text) for $^{40}$Ca($\alpha$,$\gamma$). The temperature $T$ is given in GK.}
\label{fig:ca40}\end{center}
\end{figure}

\begin{figure}[htp]
\begin{center}
\epsfig{file=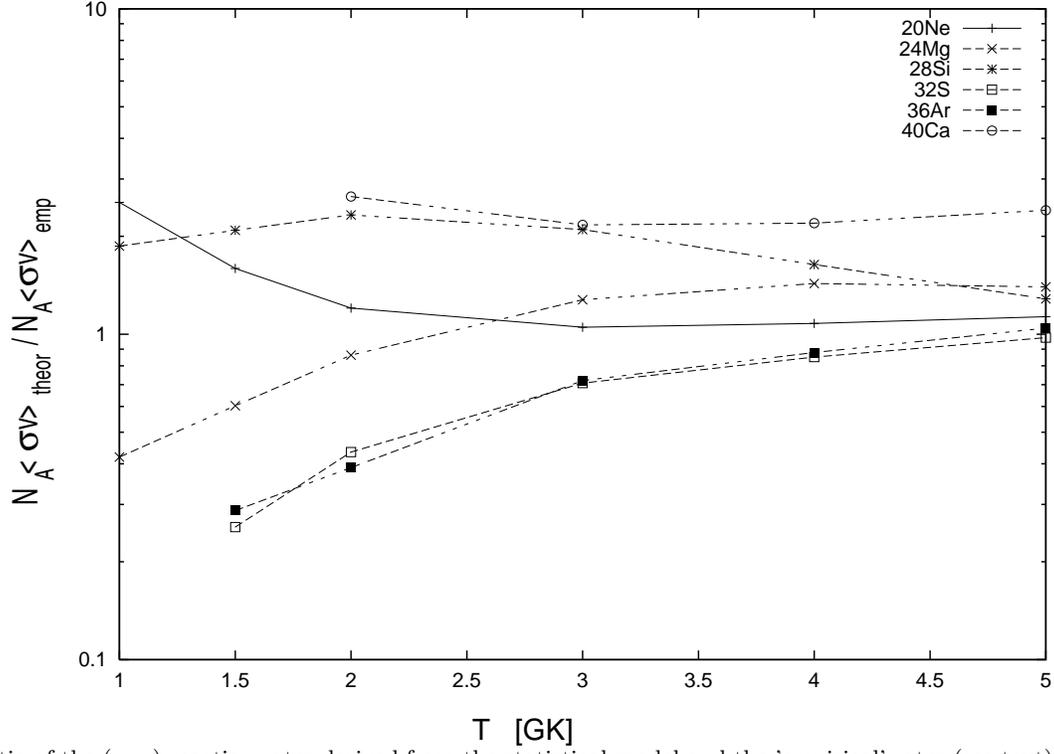,width=10cm,angle=-90}
\caption{Ratio of the ($\alpha,\gamma$) reaction rates derived from the statistical model
and the 'empirical' rates (see text) for the target nuclei $^{20}$Ne, $^{24}$Mg, $^{28}$Si,
$^{32}$S, $^{36}$Ar, and $^{40}$Ca. The ratios are displayed within the astrophysically
relevant temperature range.}
\label{fig:ratio}\end{center}
\end{figure}

\end{document}